\documentclass{article}
\pdfoutput=1
\usepackage{PRIMEarxiv}
\usepackage[utf8]{inputenc} 
\usepackage[T1]{fontenc}    
\usepackage{url}            
\usepackage{booktabs}       
\usepackage{amsfonts}       
\usepackage{nicefrac}       
\usepackage{microtype}      
\usepackage{lipsum}
\usepackage{fancyhdr}       
\usepackage{graphicx}       
\usepackage{amsmath}
\usepackage{booktabs,ragged2e}
\usepackage[flushleft]{threeparttable}
\usepackage{algorithm}
\usepackage{varwidth}
\usepackage{array}
\usepackage{fixltx2e}
\usepackage{dblfloatfix}

\title{Diversity beyond density: experienced social mixing of urban streets
}

\author{
  Zhuangyuan Fan  \\
  Department of Geography \\
  The University of Hong Kong \\
  Senseable City Lab\\
  MIT\\
  \texttt{yuanzf@mit.edu} \\
   \And
  Tianyu Su \\
  Graduate School of Design \\
  Harvard University \\
  \texttt{tsu@gsd.harvard.edu} \\
  \And
  Maoran Sun \\
  Senseable City Lab \\
  MIT \\
  \texttt{maoransu@mit.edu} \\
  \And
  Ariel Noyman \\
  Media Lab \\
  MIT \\
  \texttt{noyman@media.mit.edu} \\
  \And
  Fan Zhang \\
  Senseable City Lab \\
  MIT \\
  \texttt{zhangfan@mit.edu} \\
    \And
  Alex `Sandy' Pentland \\
  Institute for Data, Systems, and Society \& Media Lab \\
  MIT \\
  \texttt{sandy@media.mit.edu} \\
    \And
  Esteban Moro \\
  Institute for Data, Systems, and Society \& Media Lab \\
  MIT \\
  Departamento de Matemáticas \& GISC\\
  Universidad Carlos III de Madrid\\
  \texttt{emoro@mit.edu} \\
}

\begin{document}
\maketitle

\begin{abstract}
Urban density, in the form of residents' and visitors' concentration, is long considered to foster diverse exchanges of interpersonal knowledge and skills, which are intrinsic to sustainable human settlements. However, with current urban studies primarily devoted to city and district-level analysis, we cannot unveil the elemental connection between urban density and diversity. Here we use an anonymized and privacy-enhanced mobile data set of 0.5 million opted-in users from three metropolitan areas in the U.S to show that at the scale of urban streets, density is not the only path to diversity. We represent the diversity of each street with the Experienced Social Mixing (ESM), which describes the chances of people meeting diverse income groups throughout their daily experience. We conduct multiple experiments and show that the concentration of visitors only explains 26\% of street-level ESM. However, adjacent amenities, residential diversity, and income level account for 44\% of the ESM. Moreover, using longitudinal business data, we show that streets with an increased number of food businesses have seen an increased ESM from 2016 to 2018. Lastly, although streets with more visitors are more likely to have crime, diverse streets tend to have fewer crimes. These findings suggest that cities can leverage many tools beyond density to curate a diverse and safe street experience for people.
\end{abstract}


\section*{Introduction}
Diversity is intrinsic to a sustainable, resilient, and inclusive city \cite{jacobs1961uses, jacobs1993great}. Within many forms of diversity, the diverse collection of people, socially and economically, is one of the crucial preconditions of economic urban vitality \cite{glaeser2012triumph} and creativity \cite{florida2005cities, florida2006flight, landry2012creative}. On the contrary, the segregation of people was shown to impact children's economic outcomes \cite{mayer2002economic}, widening the digital gap and hindering access to public health services \cite{kramer2009segregation}. Correspondingly, researchers across the fields of economics, sociology, urban planning, and mobility have devoted themselves to explaining the level of social mixing and segregation across space and time.

Although most understanding of vitality and mixing in our cities has been done using static, residential only, and sometimes outdated census data, recent studies of human mobility further draw our attention to the activity space in cities beyond where people live. It is clear that people do not only stay where they live but also work, travel, and relax in places other than homes. Therefore, people living in less diverse neighborhoods may still have chances to encounter people with different knowledge and experience during their daily life. Recent studies have measured how well people with different backgrounds are mixed during their daily travel activities \cite{moro2021mobility, xu2019quantifying, athey2021estimating}, online communications, and purchase activities \cite{dong_segregated_2020}. It is shown that the likelihood of people meeting diverse others is related to an individual's demographic characteristics, lifestyle, and travel habits \cite{moro2021mobility}. However, beyond an individual's behavior choices, what remains unanswered is how a city as a system could build an environment that cultivates social mixing in the long run.

Many urban theories and practices have voiced the importance of density, or the concentration of people, in leading towards more urban vitality \cite{jacobs1961uses, wirth1938urbanism, glaeser2012triumph, holden2005three}, and thus favoring a more socially mixed urban environment \cite{moroni2016urban, florida2005cities}. If we view the desired outcome as the diverse admix of human knowledge, abilities, preferences, interaction, and so forth, using density as the only tool has its limitations - city blocks with a high density of office buildings could still only see people with similar income levels and skill sets. With the relationship between density and diversity inevitably being non-linear, we should further unpack what other tools cities could leverage to curate a socially mixed urban environment.

Building on the work that creates an activity-based measure of diversity and segregation, this study is primarily concerned with the space of street sidewalks. Theoretically, street sidewalks are critical urban open spaces advocated by sociologist William H. (Holly) Whyte \cite{whyte1980social}, journalist Jane Jacobs \cite{jacobs1993great}, architect Jan Gehl \cite{gehl1971life}, and New Urbanism scholars\cite{cabrera2013can}. Practically, the United Nations Sustainable Development Goals (SDGs)'s target 11 emphasizes the vital role of urban public spaces in social and economic life. However, with most studies on social mixing conducted at the city or district level, the elemental contribution along the street network to social mixing is rarely unveiled.

To understand the vital elements leading to socially mixed street experience, we first create a measure of Experienced Social Mixing (ESM) that estimates the income aspect of mixing in cities using a large collection of micro-scale mobility data across 40 counties and three metropolitan areas in the U.S. Then, we examine what factors beyond density could further contribute to social mixing. Two main sets of factors connected with human mobility and social interactions are examined in this study. We first discuss the importance of socio-economic factors, including income and residential mixing. These variables are learned from the human mobility and segregation literature that people tend to visit places at a given income segregation level \cite{moro2021mobility}, and residential segregation correlates with experienced segregation at a city scale \cite{athey_experienced_2020}. The second set of factors describes venues along the streets and how safe the street looks. These variables resonate the city planning literature that advocates the mixed-use development \cite{jacobs1993death, calthorpe1986sustainable, grant2002mixed} and street environment safety \cite{branas_citywide_2018,naik2014streetscore, jacobs1961uses}.

We present three main results. First, conditioning on density, ESM can still be explained by adjacent neighborhoods' residential mixing, income level, and venues along the street. Density, in our measure, the number of visitors visiting a street segment at any specified time, only explains around 26\% of the model estimation, while residential mixing, income level, and venues contribute to 44\% of the model estimation.

Second, ESM measured at different hours of a day is closely related to different types of venues along a street segment, highlighting the importance of a mixed-use environment. Among the venues, food-related businesses exert the highest contribution to explaining the ESM at different times of the day. Meanwhile, when controlling for the total number of visitors, the streets with more coffee and tea venues can attract more diverse groups of people. Moreover, we found that the street segments with an increase in food-related business from 2016 to 2018 are likely to see an increase in the ESM. This longitudinal effect holds conditioning on the increase of total visitors. 

The well-being of urban dwellers is a multi-dimensional concept that goes beyond diversity and economic status and involves health, crimes, and other aspects of life. Beyond the daily experienced social mixing, a body of recent literature indicates that mobility patterns also predict crimes \cite{bogomolov2014once, zhang2021perception, bogomolov2015moves} - where residents visit is also a source of neighborhood (dis)advantage \cite{levy2020triple}. To further understand the effect of the ESM, the last part of our study analyzes the relationship between visitor volumes, ESM, and different types of crime incidents around each street segment. We show that although denser cities attract more crime incidents, conditioning the visitor volumes, street-level ESM has a negative association with crime count. 

This study highlights the importance of a high-resolution measure of social mixing as a time-dynamic urban phenomenon - streets adjacent to each other could present dramatically different levels of ESM at different times of the day. Furthermore, we illustrate how cities could leverage the open space of the street sidewalks to increase the chance for different people to meet each other and thus mitigate the existing downfall of residential segregation. Lastly, our result also shows that diverse visitor experience does not always go in parallel with high volumes of crime incidents. Large cities can leverage many policy tools beyond density to curate a diverse and safe street experience for people.

\begin{figure}[!t]
\centering
\includegraphics[width=0.95\linewidth]{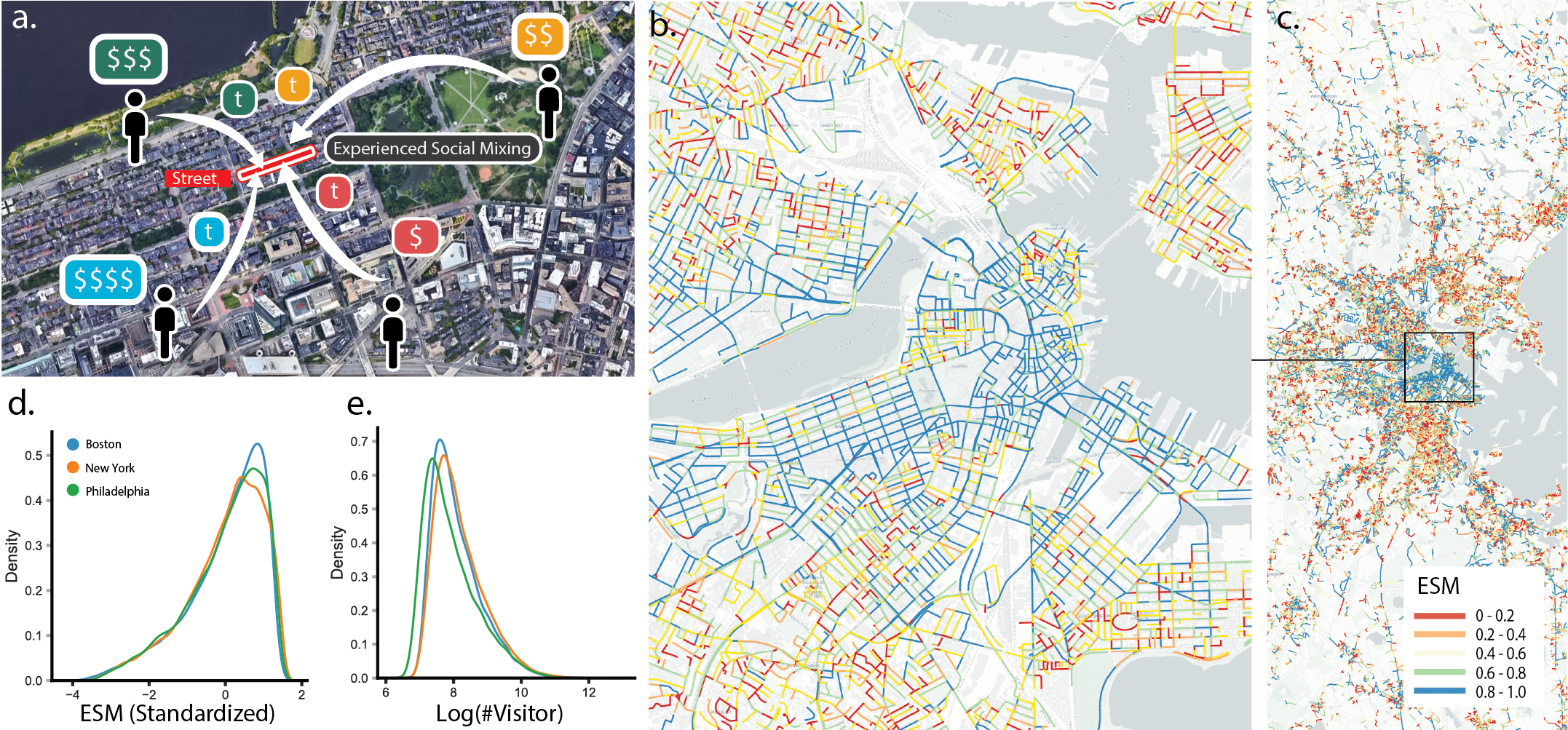}
\caption{\textbf{Measure the Experienced Social Mixing}. a. Street-level ESM is calculated based on the time spent by each income group at each street segment. b. Distribution of daily ESM of Boston. Only street segments with at least 1 POI within the 100-meter buffer area are visualized on the map. c.Regional map visualizing the ESM around Boston metropolitan area. d.Distribution of the ESM by metropolitan area. e. Distribution of log-transformed visitor counts during the study period by metropolitan area.}
\label{fig1}
\end{figure}

\section*{Results}
\label{sec:headings}
\subsection*{Street-level ESM}
We create a measure of ESM for three large metropolitan areas of Boston, New York, and Philadelphia, which involves more than 40 counties across five states (MA, NY, DL, NJ, PA). The privacy-enhanced mobility data is provided by Cuebiq, which includes 3-month long records across two years of anonymized device-level location pings for 0.5 million users who opted into data sharing for research purposes under a GDPR and CCPA-compliant framework.

To construct the ESM, we pre-process the data to identify each device's home Census Block Group (CBG) and stay locations using the same method in a previous work by Moro et al. \cite{moro2021mobility}. We first associate each device from the mobility dataset with an approximate socioeconomic status by their inferred home CBG. Each individual’s home CBG is obtained from their most commonly visited location between 10 pm to 6 am (See Methods). Then all individuals are grouped into four quantiles of income groups according to their home CBG's median household income's relation to the metropolitan area distribution of median household income (See Methods). We then extract visits an individual made to a given street segment for at least 5 minutes but a maximum of two hours. This is to prioritize sidewalk activities that have the potential for meaningful interaction among pedestrians. Activities such as visiting cafes, restaurants, parks, or simply resting along the streets are emphasized. Other activities such as working in an office building or watching movies, which usually take a long time but offer little chances for people to meet each other, are dropped. The post-stratification process reduces sample bias regarding population and income level (See Supplementary Note 2).

\begin{figure}[t!]
\centering
\includegraphics[width = 0.95\linewidth]{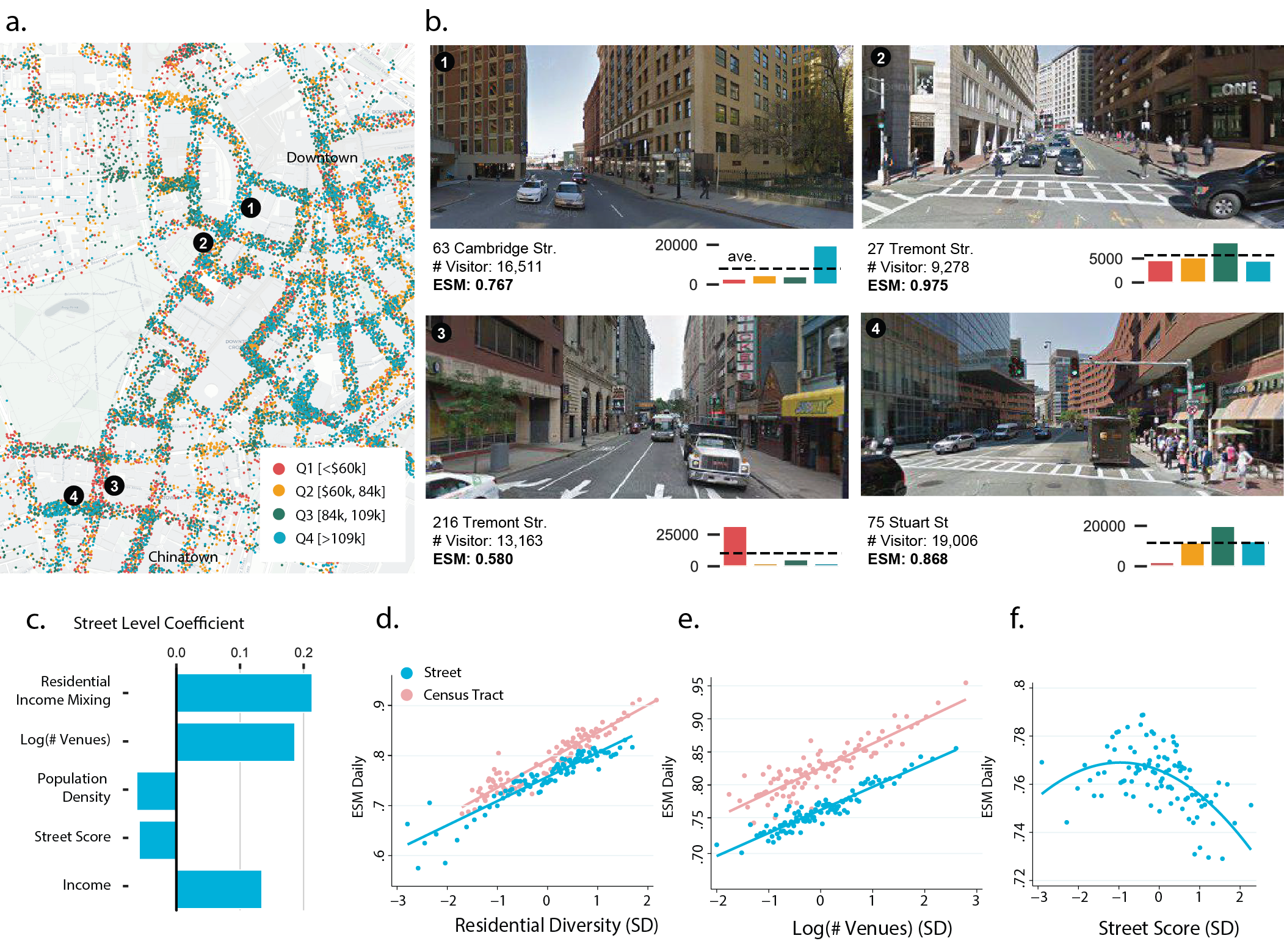}
\caption{\textbf{Explain the Experienced Social Mixing.} a. Street-level ESM representation. Each dot represents 0.5\% of time spent at a street segment by a group of visitors. b. Street view examples of streets that are adjacent to each other yet present different ESM levels. Each barplot presents the hour of visits from each income group. The total number of unique visitors is shown accordingly. c. Coefficient of all variables estimating street-level social mixing (all shown variables have p-value<0.01). d. Compare the effect of residential mixing on ESM at street and CT-level. Binned-scatter plot with 100 quantiles. e. Compare the effect of the number of POI on ESM at the street and CT-level. f. A non-linear relationship between the Street Score and the street-level ESM. (See the full table in the Supplementary Note Tab. S1)}.
\label{fig2}
\end{figure}

We compute the street-level ESM with the proportion of total time spent at that street by each income quartile (see Fig. \ref{fig1}a). The ESM is defined as the Shannon entropy \cite{shannon1948mathematical} of each income group's activity time spent at a given street segment (See Methods). ESM quantifies income mixing from 0 to 1. A street segment that is fully mixed ($ESM_s = 1$) when the total time across all individuals spent at the street segment is split evenly among the four income quartiles, while a street segment with $ESM_s = 0$ indicates that the street segment is only visited by one income group. Fig. \ref{fig2}b shows several examples. The bar chart of each street represents the accumulated time spent by each income group at the sample street. The dashed line demonstrates when the street would be fully mixed ($ESM_s = 1$). For example, 63 Cambridge Street in Boston has a lower ESM (0.767) in comparison to 27 Tremont Street (0.975), as the time spent by each income group along 27 Tremont Street is more evenly distributed. We note that there are other choices of mixing and segregation metrics. We also examine the robustness of our measure of social mixing against other metrics (See Supplementary Note 3).

Street-level ESM presents spatial heterogeneity in each city (Fig. \ref{fig1}c). However, it also has a very fine-grained spatial resolution. We illustrate this observation in Fig. \ref{fig2}a and b. Fig. \ref{fig2}a presents the proportion of time spent by each income group to each street segment in the Boston Downtown area. Each dot represents 0.5\% of time spent by each income group. Fig \ref{fig2}b demonstrates the street view samples and their associated distribution of time spent by each income group. Even two adjacent streets with the same intersection could present drastically different levels of social mixing. This finding shows that understanding vitality, density, or mixing in our cities at larger scales (e.g. Census Tracts or districts \cite{de2016death}) miss the fine-grained structure of how people interact and encounter in our cities and the relationship of the diversity of encounters and urban environment.

\subsection*{Explain the Street-level ESM}
\begin{figure}[!t]
\centering
\includegraphics[width = 0.95\textwidth]{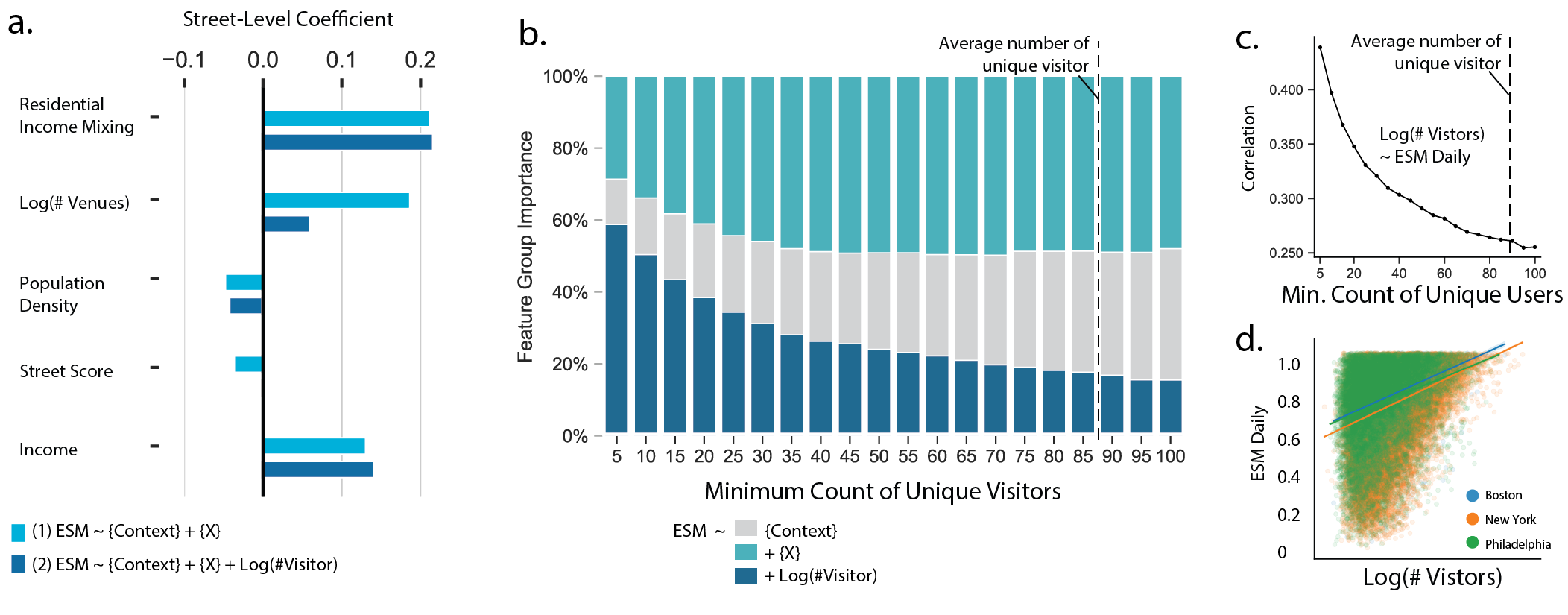}
\caption{\textbf{How much does the concentration of visitors explain social mixing?} a.Coefficients of all variables estimating the ESM (standardized) with and without the number of visitors controlled. \textit{Context} variables including the geographical fixed effects and street segment length; \textit{X} include the residential income mixing, income level, population density, the number of venues, and \textit{Street Score}. b. Summary of importance of feature groups in the regression model for social mixing by sequentially dropping street samples visited by fewer count of unique users. c.Correlation of log-transformed number of visitors and daily ESM by sequentially dropping street samples visited by fewer count of unique users. d.Relationships between the log-transformed visitor counts and daily ESM by different metropolitan areas.}
\label{fig3}
\end{figure}
To understand the relationship between the streets and ESM, we model the ESM of each street using a simple linear regression model. The explanatory variables include the street segment's length, a series of residential factors that describe the adjacent neighborhood in which the street segment is located, and the number of venues (such as restaurants, cafes, grocery stores, schools, etc.) located along the street, and how safe a street segment look. The residential factors include population density, median household income, and residential income mixing. In addition, considering the geographical differences of all samples included in the study, we also add county-level fixed effects. To compute the residential factors, we obtain a collection of CBGs that fall within a street segment's 800-meter buffer and use the median household income and population size of each CBG accordingly (See Methods and Supplementary Note 4.4). The residential income mixing mirrors the calculation of ESM. It measures the level of income mixing of the four income groups, where 0 indicates that all residents within the 800-meter buffer belong to the same income group, whereas 1 implies a fully mixed street residential neighborhood (See Methods). 

The venues on each street are obtained from Foursquares' point of interest (POI) API. A collection of 0.1 million verified venues across the study areas are used. Lastly, given that the sense of safety is one of the significant concerns determining the street activity \cite{jacobs1961uses}, we measure how safe each street looks through the \textit{Street Score}. The \textit{Street Score} of each street segment is predicted from the Google Street View images taken along the street segment. The \textit{Street Score} model was adapted from Zhang et al. \cite{zhang2018measuring}, which is a convolutional neural network trained with the data from Place Pulse \cite{salesses2012place}. We collected 1.6 million street view images through Google API across the study areas, and images taken from winter were dropped to avoid seasonal inconsistency. A similar method was used by Naik \cite{naik2014streetscore} in describing the inequality of urban safety perception.

Fig \ref{fig2}c summarizes the regression coefficients of the above-mentioned linear regression. Besides the geographical fixed effect, street segments close to a higher level of residential mixing and more venues tend to be more socially mixed. We test if this relationship holds at the different spatial units by repeating the experiment at Census Tract (CT) level. Fig.\ref{fig2}d and e plot the results in parallel. We find that Census Tracts with more mixed residential composition and venues also tend to be more mixed. 

We also find population density and \textit{Street Score} are negatively associated with the street-level EMS. The former indicates that neighborhoods with more residents do not guarantee more chances of cross-group mixing during their daily activities. To further unpack the negative association between the \textit{Street Score} and the street-level ESM, we included a quadratic term of the safety score in the same model and identify a non-linear relationship between the \textit{Street Score} and the ESM (See Fig.\ref{fig2}f). This could reflect a number of forces, including gentrification and fear of crime in cities. People tend to avoid places that look very unsafe \cite{skogan1986fear}, but in the meantime, streets with luxury settings, well-planted groves of trees, and fine furniture also indicate a sense of gentrification do not welcome social mixing \cite{kolko2007determinants}. It is worth noting that the effect of \textit{Street Score} diminishes at CT-level study, implying that the perception of street view matters more at a scale of the street segment rather than in larger spatial unit (See Supplementary Fig S3 and Table S1).

\subsection*{ESM and Density}
As many current urban theories indicate, urban density is one of the important instruments that foster diversity \cite{jacobs1961uses, wirth1938urbanism, glaeser2012triumph, holden2005three}. A street segment with more visitors could naturally have a higher chance to be more socially mixed. Could it be that the factors we discussed above are associated with the ESM through the channel of density?  

As we can see in Fig. \ref{fig3} d, the log-transformed total number of visitors visiting each street has a correlation with the street-level ESM in different cities. We find this correlation dropped as we sequentially excluded street segments with too few unique visitors to avoid the small sample bias (Fig. \ref{fig3} c). This suggests that the ESM is not solely explained by density, albeit they are correlated. Fig. \ref{fig2}b also shows specific street segment examples and implies that streets with fewer total visitors can still be more mixed than others.
\begin{figure}[!t]
\centering
\includegraphics[width=0.95\linewidth]{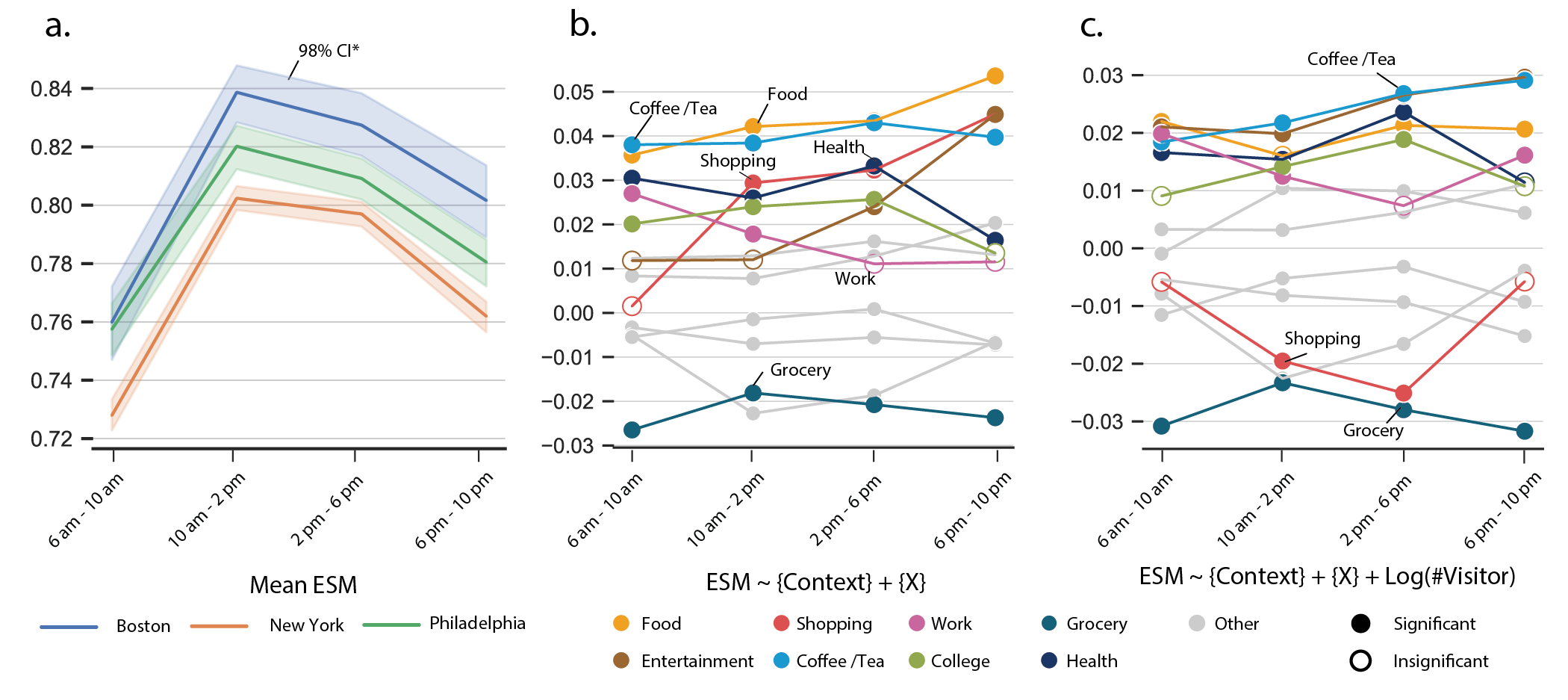}
\caption{\textbf{Time-variant ESM and street venues.} a. Mean time-variant ESM for each metropolitan area. Note that we only present the streets with at least 20 unique visitors at each time period of time. $CI^*$ stands for the confidence interval. 98\% of the confidence interval is shaded.  b. Time-variant effect of the number of venues by their categories. c. Time-variant effect of the number of venues by their categories conditioning on the visitor count. Only coefficients with $p$ value smaller than 0.05 are shown as significant.}
\label{fig4}
\end{figure}
To further quantify the factors explaining ESM beyond density, we repeat the regression model in the previous section by including the log-transformed count of visitors as a variable. Fig \ref{fig3} b shows the importance of variable groups in predicting street-level ESM. By excluding street segments with too few unique visitors (fewer than 20), the count of visitors accounts for around 26\% of the variance in street-level ESM. Apart from the geographical fixed effect and street segment length, the residential income mixing, income level, population density, and venues account for around 44\% of the model variance.

Again, Fig \ref{fig3}a plots the comparison of two regression models (See Supplementary Table S1 for the full result). We show that by including the log-transformed total visitor count as a variable, the effects of residential income mixing, income level, and population density hold. The effect of the total count of venues and the \textit{Street Score} dropped drastically. These observations lead to two insights: one is that the streets with more venues and look less safe tend to have more visitors and therefore more socially mixed. The other lies in the fact conditioning the ability to attract more visitors, street segments within areas that have more mixed residential environments and higher income levels are still more socially mixed. One might wonder if the mixed residential environment contributes to the ESM directly as a street segment's neighborhood residents would visit the street segment very often. We calculate the average travel distance from any given street to its visitors' home CBG's geometry center. We found that more than 95\% of the street segments' visitors live more than 800-meter away from the street. This result confirms that the street segment with a diverse residential environment attracts visitors from different income groups even though they don't live nearby.

\subsection*{Temporal Variation of ESM}
Street life has a unique temporal characteristic \cite{lynch1960image, jacobs1961uses}. To better understand the temporal variation of ESM throughout the day, we reconstruct the model by estimating the ESM at four selected time periods of a day. Fig. \ref{fig4}a shows the average street-level ESM for each metropolitan area throughout the day. We observe that on average, the ESM is highest around noon and lowest in the morning, reflecting the daily activities in cities. In addition, we group venues by their categories to test how the category of venues might predict ESM dynamically (See Supplementary Table S5. for venue summary by types). Similarly, we compare the model results by excluding and including the count of visitors at different times correspondingly (Fig. \ref{fig4}).

Fig. \ref{fig4}b shows that streets with more venues such as food, coffee, and tea are more mixed throughout the day, while the streets with more shopping and entertainment (bars and clubs) venues become more mixed in the late afternoon. Streets with more health-related venues tend to be more mixed before 6 pm, and a similar trend is also seen for streets with more work-related venues. We interpret these results to be associated with both the schedule of the business and people's mobility patterns. While people mostly visit hospitals and clinics during the day, streets with these venues tend to be more mixed during their normal operating hours. Since people will more likely to visit bars and go shopping after work, streets with these venues only start to see mixed groups of people later in the afternoon. We also find that streets with more grocery stores tend to have consistently lower levels of mixing throughout the day. This is likely an effect of segregated residential neighborhoods, where people tend to go to grocery stores closer to where they live. 

In parallel, we also add the number of visitors to each street segment to test if the impact of density would change the model results. Fig \ref{fig4}c shows that conditioning on the count of visitors, number of coffee, tea, food, entertainment, and health care venues are still contributing to more mixed streets throughout the day, while streets with more grocery stores are still less diverse. We understand that health care facilities are naturally more integrated given their unique service role in the city. However, the effect of coffee, tea, food, entertainment, and grocery stores reveals that even on streets with a similar amount of visitors, their level of social mixing can still vary considering the different functions it provides. After controlling the density, we also found the effect of shopping venues dropped and even reversed. We interpret this observation as that shopping venues are more successful in bringing in a high volume of people, yet the people group attracted to these areas might not be as diverse.

\subsection*{Changes of ESM from 2016 to 2018}
The cross-sectional study above highlights the food-related venues in predicting street-level ESM. We further design an experiment to test if the relationship holds longitudinally. Here, we leverage a crowd-sourced dataset, Boston's Hidden Restaurant\footnote{https://www.hiddenboston.com/closings-openings.html}, contributed by local communities from the Boston region to test if the open and close of food-related businesses from 2016 to 2018 cast any impact on the changes of street-level ESM in the corresponding time, controlling for the changes of residential features (Summary Stats included in Supplementary Table S3). Understanding that streets with a very high ESM in 2016 would have less room to improve than streets that were less mixed, we control for the ESM at 2016 for all models. In addition, the model also includes the same social and geographical context features in 2016 to account for the potential trend differences (See the full result in Supplementary Table S4).

{
\def\sym#1{\ifmmode^{#1}\else\(^{#1}\)\fi}
\begin{table*}[!ht]
\begin{threeparttable}
\caption{Change of ESM 2016 - 2018 (Street-Level)}
\small
\setlength\tabcolsep{0pt} 
\begin{tabular*}{\textwidth}{@{\extracolsep{\fill}} l*{5}{c}}
\toprule
&\multicolumn{5}{c}{$\Delta$ ESM}\\
                    &\multicolumn{1}{c}{(1)}&\multicolumn{1}{c}{(2)}&\multicolumn{1}{c}{(3)}&\multicolumn{1}{c}{(4)}&\multicolumn{1}{c}{(5)}\\

\midrule

$\Delta$ Pop Den    &       0.003         &                     &       0.003         &       0.003         &       0.003\sym{*}  \\
                    &     (0.002)         &                     &     (0.002)         &     (0.002)         &     (0.002)         \\
$\Delta$ MH Income  &      -0.000         &                     &      -0.000         &      -0.000         &      -0.000         \\
                    &     (0.002)         &                     &     (0.002)         &     (0.002)         &     (0.002)         \\
$\Delta \%$ Bachelor&       0.010\sym{***}&                     &       0.010\sym{***}&       0.010\sym{***}&       0.009\sym{***}\\
                    &     (0.002)         &                     &     (0.002)         &     (0.002)         &     (0.002)         \\
$\Delta$ Resi. Diversity&      -0.001         &                     &      -0.001         &      -0.001         &      -0.001         \\
                    &     (0.002)         &                     &     (0.002)         &     (0.002)         &     (0.002)         \\

\midrule
$\Delta$ Food Business      &                     &       0.004\sym{**} &       0.004\sym{**} &       0.006\sym{***}&       0.003\sym{**} \\
                    &                     &     (0.001)         &     (0.001)         &     (0.002)         &     (0.001)         \\

$\Delta$ Food Business
$\times$ ESM. 2016&                     &                     &                     &      -0.006\sym{**} &                     \\
                    &                     &                     &                     &     (0.003)         &                     \\
$\Delta$ Log(Visitors)&                     &                     &                     &                     &       0.017\sym{***}\\
                    &                     &                     &                     &                     &     (0.002)         \\

\midrule
Observations        &        3768         &        3768         &        3768         &        3768         &        3768         \\
R-squared           &      0.5480         &      0.5450         &      0.5484         &      0.5490         &      0.5579         \\
Fixed effect
(county) & Yes & Yes & Yes & Yes & Yes \\
\midrule
\textit{Trend Control} \\
Log($\#$ POI) & Yes & Yes & Yes & Yes & Yes\\
Resi. Controls & Yes & Yes & Yes & Yes& Yes\\
Seg. Length & Yes & Yes & Yes & Yes & Yes\\
ESM 2016 & Yes & Yes & Yes & Yes & Yes\\

\bottomrule
\end{tabular*}
\label{tab1}
\smallskip
\begin{tablenotes}
\item OLS estimates on change of ESM from 2016 to 2018. Only streets with at least 20 unique visitors in both observation periods (84 days in each year) are included. Standard errors in parentheses. $\#$ refers to count. Each street segment contains at least 3 POIs in the year 2016.

\item $^{***}$ denotes a coefficient significant at the ${0.5\%}$ level, $^{**}$ at the ${5\%}$ level, and $^{*}$ at the ${10\%}$ level.
\end{tablenotes}
\end{threeparttable}
\end{table*}
}

Table \ref{tab1} illustrates the results. Column 1 indicates that among all residential variables, the change of proportion of residents with at least a bachelor's degree is the only feature that contributes to the change in ESM. With all other features controlled, we found little relationship between the changes in residential income diversity and changes of ESM. This is partly because the residential income diversity only changes very subtly between the two years.

Columns 2 and 3 indicate that the absolute increase of the number of food businesses positively correlates with the change of ESM. The coefficient of change of education level still holds by including the change of food business. Column 4 includes an interaction term to test the marginal effect of the food business, considering that streets with different original ESM in 2016 might respond to the changes differently. We show that the interaction term of ESM 2016$\times \Delta$ food businesses is negatively associated with the change of ESM. It implies that with a similar increase of food businesses, the streets with a lower ESM in 2016 tend to see more increase of ESM (Fig \ref{fig4}a). Fig \ref{fig4}b shows examples of streets with changes in the food business and different levels of ESM in 2016. Lastly, we further include the log-transformed changes of visitors to each street segment between 2016 to 2018 in Column 5. As expected, the increase in the number of visitors contributes to the increase of ESM. However, the effects of food business and education level still hold with the changes in visitor volumes.

We also tested the effect of newly established businesses using a different data source (Reference USA) and \textit{Street Score} on the changes of ESM. Supplementary Table S4 reports all results. The change of \textit{Street Score} does not have a significant connection with the change of ESM. This is also potentially due to the fact that the change in urban appearance between the two years is relatively subtle. Consistent with the result of the change of the food business, the establishment of new business has a positive effect on the change of ESM.

\subsection*{ESM and Crime}
\begin{figure}[!t]
    \centering
    \includegraphics[width = \textwidth]{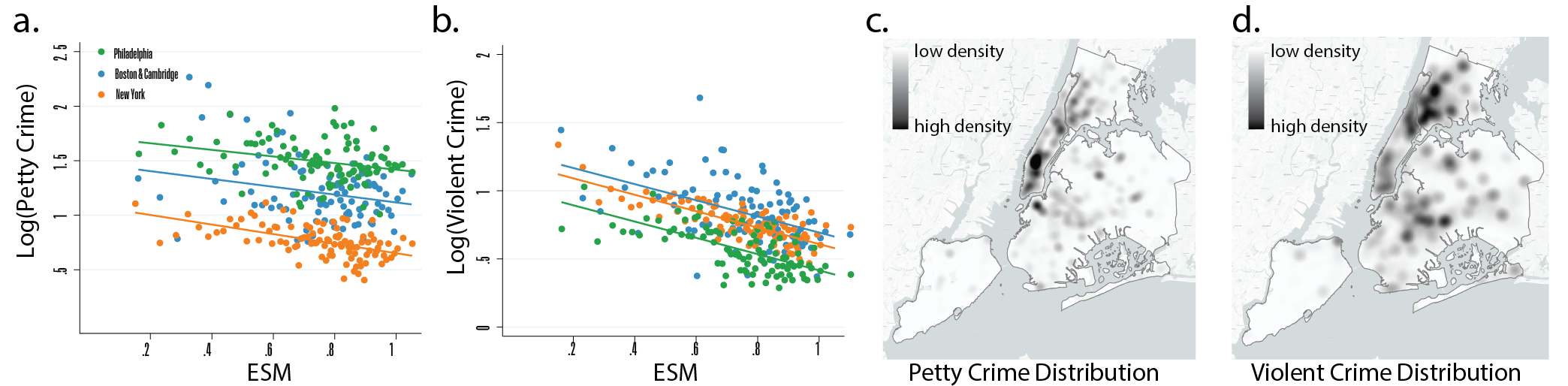}
    \caption{\textbf{ESM and Crime.} a. Binned scatter plot between ESM and log-transformed number of petty crimes (with number of visitors by street segment controlled). b. Binned scatter plot between ESM and log-transformed number of violent crimes (with number of visitors by street segment controlled). c. Kernel density illustration of petty crime density in New York City. d. Kernel density illustration of violent crime density in New York City. }
    \label{figcrime}
\end{figure}

One of the main concerns with dense cities is crime \cite{moroni2016urban, glaeser1999crime}. If a higher ESM indicates a higher chance for people with diverse backgrounds to meet each other, will it lead to more crime incidents? To understand the potential connection, we obtain crime reports from four sample cities within our study areas: New York City, Boston, Cambridge, and Philadelphia. Fig \ref{figcrime} a and b plot the relationships between crimes and the street-level ESM, conditioning on the number of visitors. We found that the number of petty and violent crimes (see Method for detailed definition) has a negative relationship with ESM. Conditioning on residential population, income level, number of visitors, residential diversity, and the number of POI, the street segments with one standard deviation higher ESM is associated with 2.6\% fewer violent crimes and 2.4\% fewer pretty crime (See Supplemental Note 8 for the full results). This result implies that social mixing does not need to come at the price of more crimes. On the contrary, we can still create a socially mixed street environment with fewer crimes. 

\section*{Discussion}

Many forces like gentrification, redlining, housing, or income inequality tend to segregate people in our cities. Curating a socially mixed urban environment is a common challenge for cities that aim at the overall goal of sustainability. Our study contributes to the current literature in this context from three perspectives. First, we investigate how the street sidewalk, as one of the most significant urban public spaces, can bring people from different income backgrounds together. The social mixing measured at street segment level rather than neighborhood or city represents the direct environment people will encounter through their day of life in cities. A street segment's immediate areas' residential mixing and the number of venues it is adjacent to have a strong connection with how socially mixed it is. Second, we show that social mixing can exist beyond density. Street segments with a similar number of visitors could still present different levels of social mixing, and the remaining variance could largely be explained by the residential mixing and income level. In addition, by further modeling the level of social mixing at different periods, we also present its temporal trend that is connected to the types of venues a street contains. We demonstrate that streets with more food, cafe, and tea places are still more mixed conditioning on the number of total visitors, while the streets with more grocery stores tend to be less mixed. Lastly, we found that conditioning on the number of visitors, the ESM has a negative association with street-level crime count. 

Our results suggest venues for city governments and planning agencies to foster multifarious hubs for their citizens through the street life they may experience. While the street life in cities has mostly been described in planning literature through ``vitality'' \cite{jacobs1993death, de2016death}, we show that high density, or the large volume of people, is not the only path towards diversity. The increase in education level and establishment of some type of business (e.g., food-related) around a street can further bring more people with diverse income backgrounds together.

In addition, building upon the literature that has incorporated the mobility pattern into crime prediction, we highlight that socially mixed street environments are less likely to see violent or petty crimes compared to the streets with similar visitor volumes but are less mixed. This result implies future research to test the value of mixed-use development in crime reduction.

Our study has several limitations. First, this study only discusses social mixing from the aspect of income. Other forms of social mixing, measured from race occupation, might have a different presentation from income mixing. Our measure of social mixing uses proximity as a proxy, thus cannot determine if people staying at the same street segment are having meaningful interaction or not.

\section*{Methods}

\subsection*{Street Segment}
All street segments are downloaded from the OpenStreetMap through the python omnsx package. Each street segment is defined as a segment of a street between two intersections. Each street segment contains an ID from the OpenStreetMap, a pair of u, v values representing the intersection node, and the function type of the street segment. Duplicated geometries are removed from the original dataset. We also remove the major highway, primary link, secondary links, trunks, services streets, footpaths, steps, and slopes from the original dataset. (Noted here that although we focus on the pedestrian network, footpaths are removed from the original dataset given its complexity and potential duplicates in a small space.) Only street segments with at least one POI within a 100-meter buffer radius are included in the study. A total of 151,680 street segments from three cities are included in the study.

\subsection*{Mobility Data}
\subsubsection*{Attribution of stays to streets}
Each street segment is represented as a line in space. To attribute each stay to a street segment, we find the closest street segment for each stay using the python geopandas package. To avoid attributing a stay to a distant street segment, we choose only the street segment within a $d_{max} = 100$ meters from each stay. If a stay is further than $d_{max}$ from any street segment, we discard it from the dataset. 50\% of stays are within 26.7-meter radius from their closest street segment. The average distance of a stay to the closest street segment is around 31.2 meters for all three cities. Distance is calculated based on each state's NAD83 state plane projection\footnote{New york: EPSG:2263; Boston: EPSG:2250; Philadelphia: EPSG:2272}.

\begin{table*}[!t]
\centering
\small
\caption{Summary statistics of the dataset per metropolitan area}
\begin{tabular}{@{}lllllll@{}}
\toprule
Metropolitan Area              & \# Devices & \# Stays & \# Street Segment & \# Census Tract & \# POIs & \# Street Images \\ \midrule
Boston-Cambridge-Newton (2016) & 66k       & 6.3M  & 17k             & 1007            & 14.4k   & 0.23M          \\
Boston-Cambridge-Newton (2018) & 174k       & 12.2M  & 17k             & 1007            & -   & 0.18M         \\
New York-Jersey City           & 210k      & 42.2M & 77k            & 4682            & 64.5k  & 0.41M          \\
Philadelphia-Camden-Wilmington & 112k      & 14.8M & 26k             & 1477            & 19k  & 0.75M          \\ \bottomrule
\end{tabular}
\label{tabsummary}
\smallskip
\scriptsize
\begin{tablenotes}
\item Statistics summary for three cities included. Number of devices only include those that are identified with a home census block group. Number of Stays only include the stays that are not within 50-meter buffer of their associated home location.
\end{tablenotes}
\end{table*}

\subsubsection*{Street Level Activities}
As we only focus on street-level activities, all stays that have a longer duration than 2 hours are discarded from the dataset. Any stays within a $d_{home} = 50$ meters from the identified home locations are also dropped from the study. A total number of stays and unique devices are shown in Table \ref{tabsummary}.

\subsubsection*{Identifying home and economic status}
For each smartphone, we use its stays from 22:00 to 6:00 and spatially cluster them using the Density-based spatial clustering of applications with noise (DBSCAN) \cite{ester1996density} algorithm to detect the most likely cluster of stays each individual is located in during nighttime and early morning hours. We use 2 as the minimum number of points per cluster and $\epsilon$ = 50 meters as the neighborhood. Then we join all detected cluster centers with each CBG geometry. We only consider individuals who were at the same CBG geometry for more than 5 nights in the observation period (three months), and this CBG is considered as the home for this user. We use this CBG's median household income during the associated year to estimate the user's income level. This process leaves us to consider only 0.5 million users. Post-stratification was implemented to assure the representatives of the data in terms of income and population.

\subsection*{Measuring ESM}
\subsubsection*{Create Income Groups}
We compare the median household income inferred from each individual's home CBG with the distribution of income in the metropolitan area so each CBG is assigned to a quartile of economic status within each metropolitan area. For each metropolitan area, the intervals of median household income for each economic group are different (See Supplementary Note Fig. S1)
\subsubsection*{Street-level Measure}
To measure the ESM of each street $s$ in each city, we compute the proportion of total time spent at that street $s$ by each income quartile $q$ during the selected period $h$, $\tau_{qsh}$. Then we define $ESM_{hs}$ as the Shannon entropy\cite{shannon1948mathematical} of each income group’s activities at a given street segment $s$'s during a given time frame $h$:
\begin{equation}
ESM_{sh} = -\frac{1}{log4}\sum_{q=1}^{4}\tau_{qsh}log(\tau_{qsh}),
\end{equation}
where $ESM_{hs}$ equals 0 when all users visit the street $s$ in time period $h$ are from the same income group, while a larger value of the $ESM_{hs}$ means users from all four income groups spend a more equal amount of time visiting the street $s$ during period $h$. Only street segments with at least 20 users during a given period $h$ are included in the study to avoid severe small-sample bias. The daily ESM only considers stays from 6 am to 10 pm. The ESM at other periods is as specified in the paper.

\subsubsection*{Census Tract-Level Measure}
Like street-level ESM, the CT level ESM is the entropy of each income group's activities within a CT during a given time frame. Each stay is attributed to a CT through a spatial join process. 

\subsection*{POI data}
\subsubsection*{POI to Street}
POI data is from Foursquare. We assign each POI to a street segment if it falls into a street segment's 100-meter buffer. Each POI is also joint spatially with a CT that it falls into. The POI distribution within each city is shown in Supplementary Note Table S5.

\subsubsection*{Change of Business}
The change in the food business is obtained from Boston's Hidden Restaurant website. The data contains the restaurants, cafes, bars, and other food-related businesses that are closed or open in each month since 2007. For this study, as the mobility data covers October to December in 2016 and October to December in 2018, we only select the restaurants that are either open or closed from Jan. 2017 to September 2018. The latitude and longitude of each restaurant were verified through google geocoding API. The chain stores is verified through Yelp. For the open and closed months for each recorded store, we also found similar results through the date of yelp reviews.

\subsection*{Street Score}
To quantify the physical appearance of the built environment, we obtain 360 degrees panorama Google Street View (GSV) images of streetscapes through Google Maps API in all three study areas. Each panorama is associated with a unique identifier, latitude, longitude, month, and year of when the image was captured. We specify four angles to capture the full panorama of each street view location. To avoid the seasonal effect, we only keep images taken between April and October. GSVs taken from 2015, and 2016 are used in the cross-sectional study for the 2016 panel. GSV's taken from 2018 and 2019 were used for the 2018 panel. Moreover, images that were taken interior or highway only were excluded from the dataset. A total number of 1.5 million GSVs were used in the study (Table \ref{tabsummary}).

We measure the appearance of the built environment with a ``Street Score,'' which indicates the perception of the safety of a GSV image. We use a Deep Learning model \cite{zhang2018measuring} pre-trained with a crowd-sourced dataset called Place Pulse, which contains millions of ratings on around 110,000 street view images from all over the world \cite{naik2014streetscore}. The image diversity and rating consistency were evaluated by previous works \cite{naik2014streetscore, dubey2016deep}, indicating no significant bias depends on raters' cultural backgrounds in the dataset. We predicted the perception of safety for each image by ignoring the features of the sky, cars, and people in the dataset to minimize effects from time of day and other dynamic events (See Supplementary Note 6). The predicted continuous score ranges from 0 to 10, with 0 being the least safe-looking and 10 being the most safe-looking view. Then the ``Street Score'' for each CT and street segment is the average score of all images associated with the CT and the street segment. 

\subsection*{Other Data}
Demographic data at the level of CBG and CT were obtained from the 5-year American Community Survey ACS (2012-2016, 2014-2018).

\subsubsection*{Residential Income Mixing}
 The residential income mixing is calculated using the same income group quartile per metropolitan area. To be consistent with the ESM calculation, at street level, we first buffer the street for 800 meters and extract all CBGs that intersect with the street buffer. Then using the pre-assigned income quartile based on each CBG's median household income and population, we calculate the street-level residential mixing as the equation below:
\begin{equation}
R_{c} = -\frac{1}{log4}\sum_{q=1}^{4}n_{qc}log(n_{qc}),
\end{equation}
where $n_{qc}$ is the population with the median household income level belonging to income quartile $q$. To test the robustness of this method, we also repeat the calculation by buffering from the street at 400 and 1000 meters (See Supplementary Note 4.4).

\subsubsection*{Crime Incidents}
The 2016 crime reports within the four sample cities are downloaded from each city's open data website. The original crime data comes with crime primary types, crime incident date, and address. All four cities also provide crime incident locations' latitudes and longitudes except Cambridge, MA. We retrieve the latitude and longitude of crimes in Cambridge using the Google Map API geocoding service. Then we aggregated each crime incident to the street level by associated crimes to a street segment within a 30-meter buffer distance. Two main types of crimes are separated from the original data. Violent crimes include rape, robbery, felony or aggravated assault, and homicide or murder. Petty crimes include theft and larceny. 

\subsubsection*{Other}
The population density, percentage of people with at least a bachelor degree, and median household income are derived from the same CBGs to calculate the residential income diversity. 

\subsection*{Regression Specification}
\subsubsection*{Cross-sectional Model}
We specify the ordinary least square model to explain the ESM at street level and CT level.

\begin{equation}
\label{eqols}
Y = \{Context\} + \{X\}
\end{equation}

\begin{equation}
\label{eqols2}
Y = \{Context\} + \{X\} + \{Density\},
\end{equation}
where Y is the estimated ESM of each experiment. $\{Context\}$ is a set of variables to control for the geographical context, including the segment length for the street-level experiment, land area size for the CT-level experiment, and county-level fixed effects. \{X\} includes the variables we are interested to test: residential mixing, income, population density, venues count, and \textit{Street Score} predicted from street view image data. The median household income comes from the ACS (5-year) 2011 - 2016 survey corresponding to the mobility data's associated year. To account for the effect of density, we include a $\{Density\}$ term in equation \ref{eqols2}, which stands for the total number of visitors.

\subsubsection*{Difference-in-Difference Specification}
To answer the question of which features contribute to the change of ESM, we specify the following equation:
\begin{equation}
\begin{aligned}
    \Delta Y = \{\Delta R\} + \{\Delta B\} + \{X\}\\
    + Y_{2016} \\ + Y_{2016}\times\{\Delta B\}, 
    \end{aligned}
\end{equation}
where $\Delta Y$ is the change of ESM from 2016 to 2018. $\{\Delta R\}$ is a set of demographic variables that change values from 2016 to 2018. The demographic data of 2018 is from ACS 2013 - 2018 survey. The $\{\Delta B\}$ is the change of food business aggregated at street level. $\{X\}$ includes a set of demographic variables in 2016 to control for the trend. To test the robustness, we also used the number of newly established businesses from the Reference USA 2017 data as the $\{\Delta B\}$ in additional tests.

\section*{Acknowledgements}
We would like to thank Cuebiq who kindly provided us with the mobility data set for this research through their Data for Good program. E.M. acknowledges support by Ministerio de Ciencia e Innovación/Agencia Española de Investigación (MCIN/AEI/10.13039/501100011033) through grant PID2019-106811GB-C32.
x
\section*{Author contributions statement}
Z.F and E.M. designed the research; Z.F. performed the analysis; T.S., M.S., F.Z. performed part of the analysis; Z.F. wrote the original manuscript; E.M, A.N., T.S., M.S., and F.Z. revised the manuscript; A.P. edited the manuscript. All authors reviewed the manuscript.

\section*{Code Availability}
The analysis was conducted using Python and Stata. Code to reproduce the main results in the figures from the aggregated data is publicly available on GitHub: https://github.com/brookefzy/social-mixing-street.

\section*{Competing Interests}
The authors declare no competing interests.

\bibliographystyle{unsrt}  

\end{document}


\maketitle

\tableofcontents
\newpage

\listoffigures

\listoftables

\newpage

\setcounter{figure}{0}
\setcounter{table}{0}

\maketitle

\section{Location Data}
\begin{figure*}[!htb]
    \centering
    \includegraphics[width =0.95\linewidth]{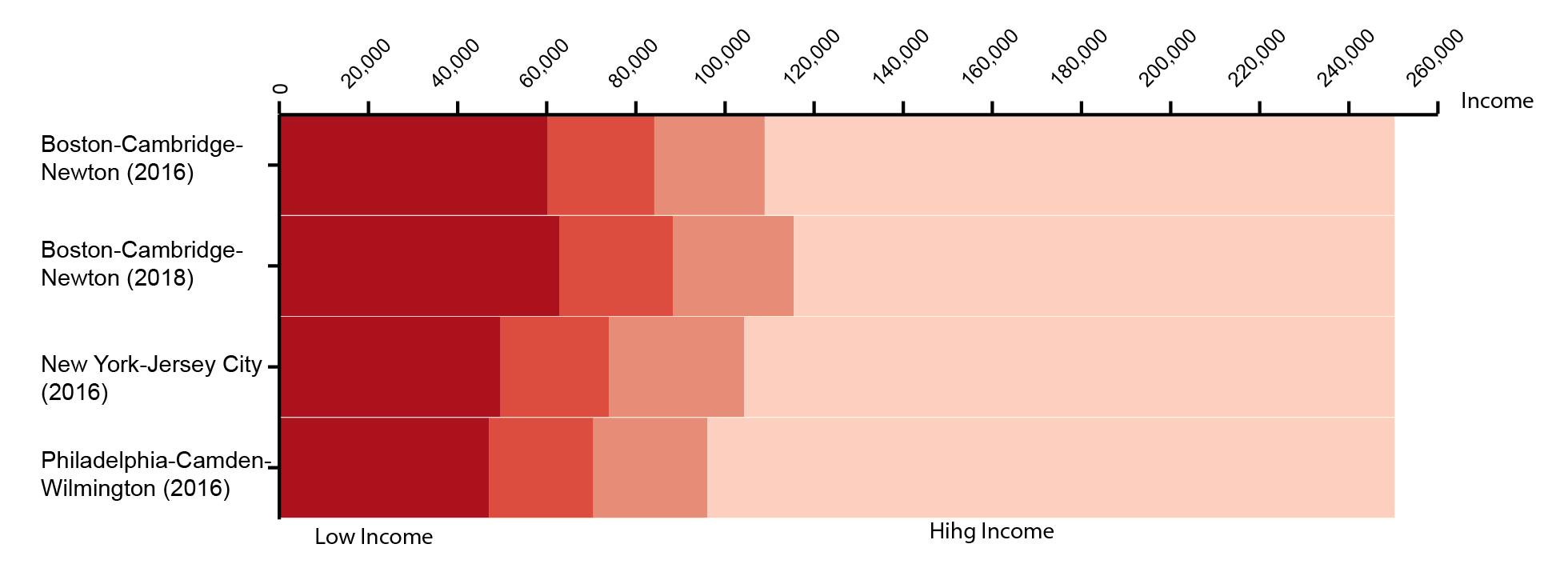}
    \caption{Income quartile distribution by each metropolitan area.}
    \label{sfig1}
\end{figure*}

\subsection{Home Location and Income Level}
We compare the median household income inferred from each user's home CBG with the distribution of income in the metropolitan area so each area is assigned to a quartile of economic status within each metropolitan area. Users in our dataset are grouped in equally-sized income quartiles in the associated metropolitan area. Intervals of median household income for each economic group are shown in Fig \ref{sfig1}.

We also test the robustness of our income group assignment using CBG's median household income. In each CBG, we can obtain a distribution of household income. We reassign all user per income distribution within their home CBG. After that, we group the individuals into four quartiles and recompute the ESM per street and CT level. We run 100 realizations of this stochastic income assignment and compare the original ESM with the average of those 100 realizations. We found that the street and CT-level ESM are correlated with the original ones ($\rho = 0.815$ for streets and $\rho = 0.814$ for CTs)

\section{Representativeness of the data}
One limitation of this dataset lies in the fact that we only observe a sample of individuals rather than the full population. So we use a weighting mechanism (post-stratification) \cite{salganik2019bit} based on the ratio of smartphone users to the true population in the CBG. We define the weight $\lambda_b$ for the individual user from block group $b$ to be:
\begin{equation}
\lambda_b = \frac{N_{b}}{\tilde{N}_{b}},
\label{eqweight}
\end{equation}
where $N_{b_i}$ is the population of CBG $b$ and $\tilde{N}_{b}$is the number of smart phone users in our sample with home locations in block group $b$. The weight $\lambda_b$ was set to 0 when a CBG has fewer than 7 unique user from our dataset to avoid over-weighting.

Then we can weight the time people from block group $b$ spends at each street $s$ and CT $c$ by
\begin{equation}
{\tilde\tau_{bs}} = \lambda_b\tau_{bs},
\label{eqweight2str}
\end{equation}
\begin{equation}
    \tilde{\tau_{bc}} = \lambda_b\tau_{bc},
    \label{eqweight2ct}
\end{equation}
where we assume that $\tau_{bs}$ is proportional to the number of people visiting the street and CT. Then we can recompute the amount of time people of quartile $q$ going to a street $s$ and CT $c$ as
\begin{equation}
    \tilde{\tau_{qs}} = \sum\tilde{\tau_{bs}} = \sum\lambda_b\tau_{bs},
\end{equation}

\begin{equation}
    \tilde{\tau_{qc}} = \sum\tilde{\tau_{bc}} = \sum\lambda_b\tau_{bc}.
\end{equation}
Fig \ref{sfigrobust1}a shows how street-level ESM changes when the weighted time $\tilde{\tau_{qs}}$ is used instead of the unweighted $\tau_{qs}$. As we can see the results have a Pearson's correlation of around 0.92.

\section{Other metrics for Experienced diversity}
\begin{figure*}[!htb]
    \centering
    \includegraphics[width =0.9\linewidth]{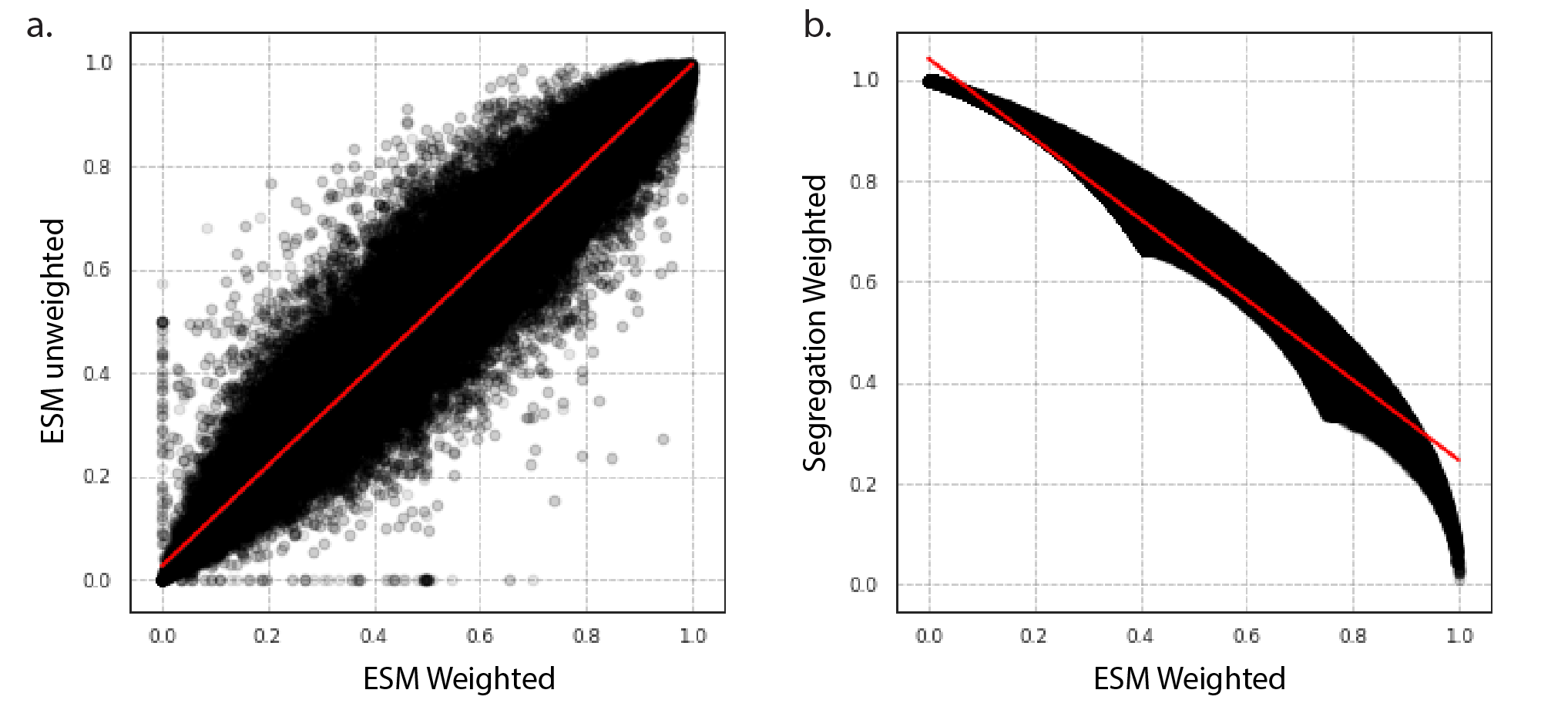}
    \caption{a. Scatter plot of user ESM using the weighted and unweighted time spent at each street segment. b.Correlation between the ESM and Segregation Index. }
    \label{sfigrobust1}
\end{figure*}

Another possible metric to measure the unevenness of the distribution of groups is the segregation index\cite{moro2021mobility}:
\begin{equation}
    S_{sh} = \frac{2}{3}\sum_q|\tau_qsh -\frac{1}{4}|
\end{equation}
where $S_{sh}$ equals to 1 when a spatial unit is fully segregated, meaning it is only visited by one individual income group. $S_{sh}$ equals to 0 when the time visited to the spatial unit is evenly distributed among all four quartile income groups. Fig. \ref{sfigrobust1}b shows the correlation plot between the segregation index and the ESM using all three cities' data in 2016. 

\section{Explain the ESM}
\subsection{Total ESM}
We first explain the ESM with residential variables, the build environment variables using an OLS model. The result is shown in the Table \ref{subtab1}. The table reflects result on streets with at least 5 POIs, 9 GSVs, total unique users above 20, adjacent CBGs with at least 50 population. CTs are selected with at least 100 GSVs,  population over 100, total unique users above 100 for the study period. VIF values for all variables range from 1 to 4.1.

{
\begin{table*}[htbp]\centering
\def\sym#1{\ifmmode^{#1}\else\(^{#1}\)\fi}
\small
\caption{Explain the ESM}
\begin{tabular}{l*{8}{c}}
\toprule

                    &        \multicolumn{4}{c}{Street} & \multicolumn{2}{c}{Census-Tract}        \\
                    \cmidrule{2-5}\cmidrule{6-7}
                     &\multicolumn{1}{c}{Log(Visitors)}&\multicolumn{3}{c}{ESM}&\multicolumn{2}{c}{ESM}\\
                     \cmidrule{3-5}\cmidrule{6-7}

                    & (1) & (2) & (3) & (4) & (5) & (6) \\
\midrule

Street Length       &       0.261\sym{***}&       0.007         &       0.007         &      -0.076\sym{***}\\
                    &     (0.007)         &     (0.006)         &     (0.006)         &     (0.006)         \\
Land Area    &                     &                     &                                        &  &      -0.116\sym{***}&      -0.113\sym{***}\\
                     &                     &                                                               &  &  &     (0.033)         &     (0.034)         \\
Log(Pop. Den)       &      -0.012         &      -0.066\sym{***}&      -0.063\sym{***}&      -0.062\sym{***}&      -0.083\sym{**} &      -0.029         \\
                    &     (0.011)         &     (0.012)         &     (0.012)         &     (0.011)         &     (0.037)         &     (0.037)         \\
Log(MH Income)      &      -0.034\sym{***}&       0.117\sym{***}&       0.118\sym{***}&       0.128\sym{***}&       0.095\sym{***}&       0.106\sym{***}\\
                    &     (0.007)         &     (0.008)         &     (0.009)         &     (0.008)         &     (0.021)         &     (0.021)         \\
Resi. Mixing         &      -0.025\sym{***}&       0.200\sym{***}&       0.201\sym{***}&       0.208\sym{***}&       0.345\sym{***}&       0.337\sym{***}\\
                      &     (0.006)         &     (0.006)         &     (0.006)         &     (0.006)          &     (0.014)         &     (0.014)         \\
Log(\#POI)           &       0.426\sym{***}&       0.195\sym{***}&       0.194\sym{***}&       0.060\sym{***}  & 0.300\sym{***}&       0.107\sym{***}\\
                     &     (0.007)         &     (0.006)         &     (0.006)         &     (0.007)        &     (0.014)         &     (0.022)         \\

Street Score         &      -0.119\sym{***}&      -0.039\sym{***}&       0.608\sym{**} &      -0.001      &       -0.026         &      -0.001         \\
                      &     (0.006)         &     (0.006)         &     (0.226)         &     (0.006)       &     (0.016)         &     (0.016)         \\
Street Score (sq)&                     &                     &      -0.648\sym{***}&                     \\
                    &                     &                     &     (0.225)         &                     \\

Log(Visitors)       &                     &                     &                     &       0.316\sym{***}&                    &       0.263\sym{***}\\
                   &                     &                     &                     &     (0.006)          &                   &     (0.023)         \\

\midrule
Observations        &       26080         &       26080         &       26080         &       26080          &        4002         &        4002         \\
R-squared           &      0.2587         &      0.1516         &      0.1519         &      0.2256          &      0.3463         &      0.3677         \\
County fixed effect & Yes & Yes & Yes & Yes & Yes & Yes  \\

\bottomrule
\end{tabular}

\label{subtab1}
\end{table*}
}

\begin{figure*}
    \centering
    \includegraphics{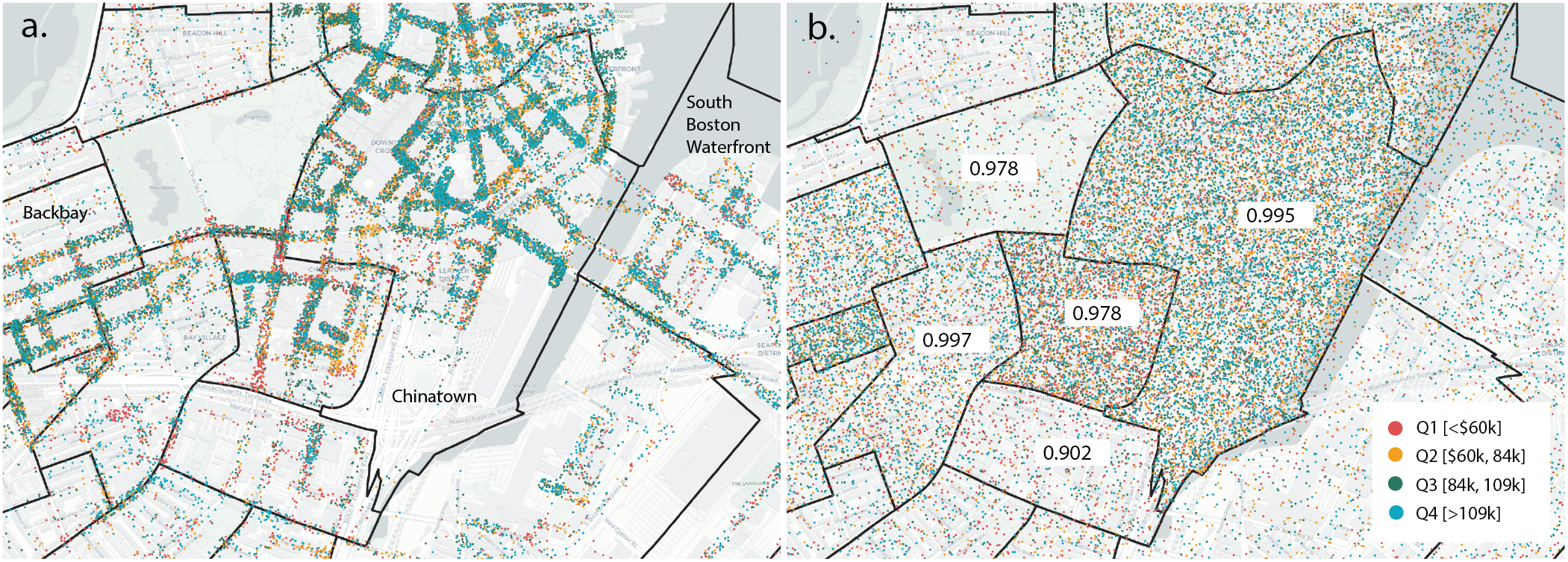}
    \caption{Street and Census Tract-Level Representation of Visiting from different groups}
    \label{subfigctstrcomp}
\end{figure*}

\subsection{Time-variant ESM}
To understand the temporal ESM, we construct the ESM per four selected time periods: 6-10 am, 10 am - 2 pm, 2 pm to 6 pm, 6 pm - 10pm. The time variant regression result is shown in Table \ref{subtabtimevariantESM}

\subsection{Density}
\begin{figure*}
    \centering
    \includegraphics{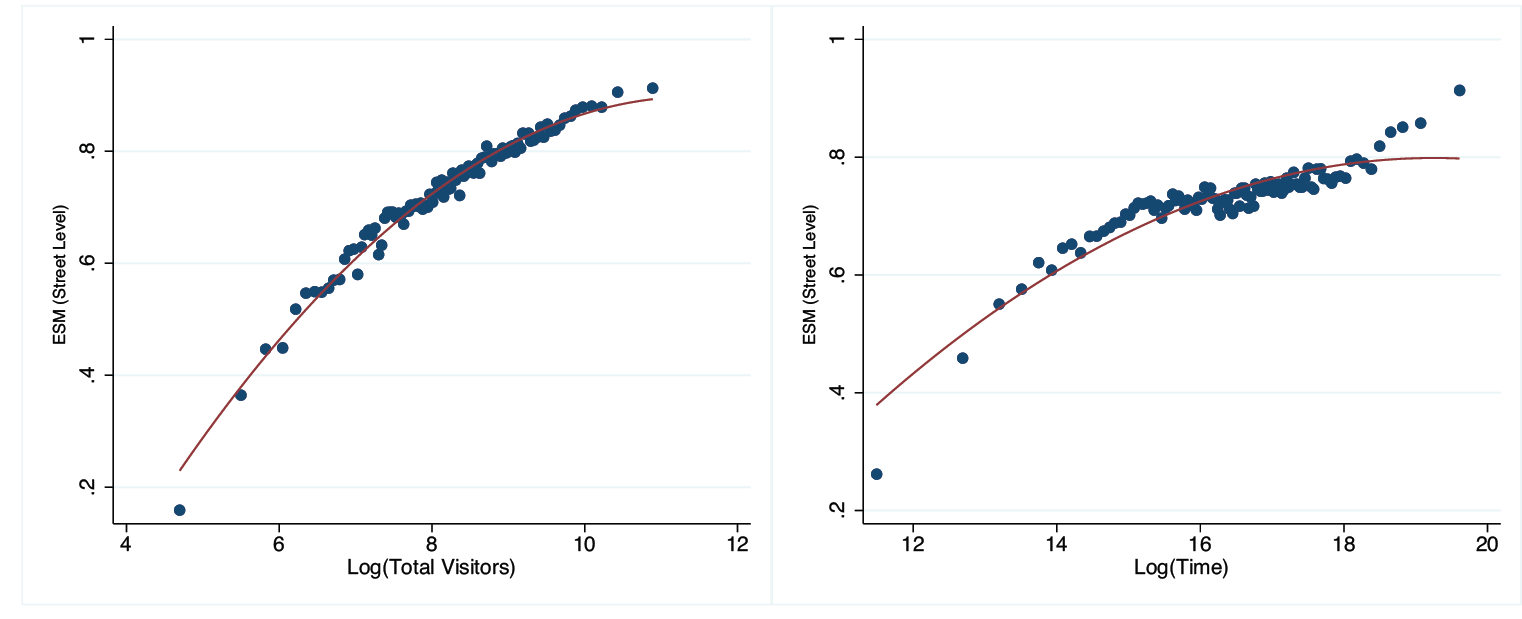}
    \caption{Binned scattered plots between the street level visiting density and the ESM. Both plots are weighted by segment length.}
    \label{subfigdensitydiversity}
\end{figure*}

We explore the correlation between the visiting density and the ESM. Understanding that the density could be one main drive of diversity, we plot the relationship between the visitor and ESM, and visiting time and ESM, respectively (Fig. \ref{subfigdensitydiversity}).

Further, we calculate the importance of feature groups after including the log-transformed visitor counts into the model. We observe that the visiting volume accounts for less than 50 \% of the model importance in all models. The relative importance of each feature group is calculated using the approach of Linderman, Merenda and Gold (LMG).

\subsection{Street-level residential variables}
Street-level residential variables are calculated using a collection of CBGs that intersect a street segments' 800 meter buffer. The population of a street segment is the total population of all intersected CBGs. The percentage of people with at least a bachelor degree is the weighted average of $\% Bachelor$ of all intersected CBGs based on their population sizes. The residential income diversity is calculated using the median household income of all intersected CBGs. We first assign each CBG with an income quartile. Then we compute the residential income diversity as:
\begin{equation}
    R_{c} = -\frac{1}{log4}\sum_{q=1}^{4}n_{qc}log(n_{qc})
\end{equation}
where $n_{qc}$ is the population with the median household income level belonging to income quartile $q$. We repeat this calculation using different size of buffer (400m, 1000m). The residential income diversity calculated with different buffer sizes are correlated. ($\rho = 0.93$ at 1000m buffer, $\rho = 0.75$ at 400m buffer.)

\subsection{Independence of residential and places variables}
\begin{figure*}[!htb]
    \centering
    \includegraphics[width =0.9\linewidth]{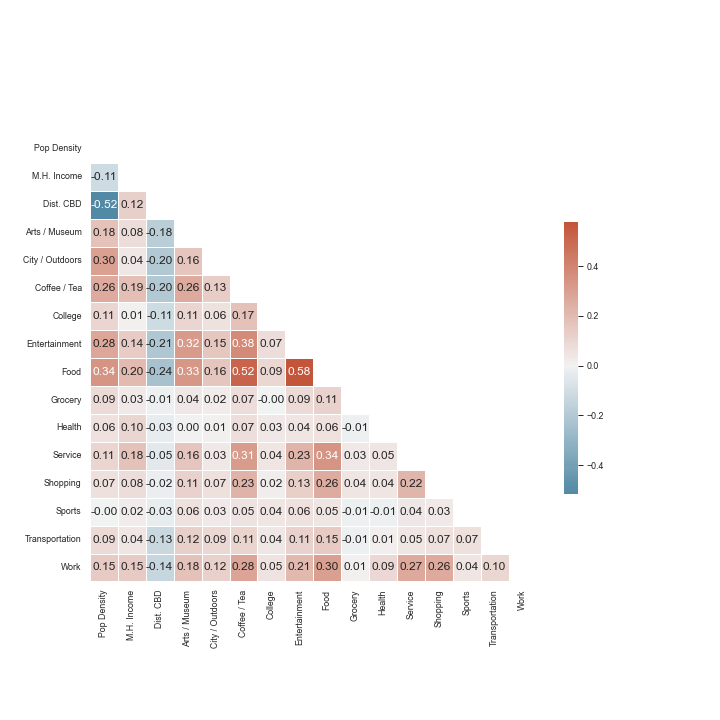}
    \caption{Coefficient matrix of all venues used in the model.}
    \label{sfigcoeff}
\end{figure*}

The Fig \ref{sfigcoeff} plot the coefficient matrix of all venues used in the study.

We also computed the variance Inflation Factor(VIF) for all variables in our models for street ESM to test for potential multicollinearity. In our models, the VIFs ranges from 1 to 4 (mean VIF at 2.2) indicating no significant issue of multicollinearity.

{
\def\sym#1{\ifmmode^{#1}\else\(^{#1}\)\fi}
\begin{table*}

    \caption{Time-variant ESM and POI Types (POI JOIN to Street 100m Buffer)}
    \centering
    \scriptsize

    \begin{tabular}{l*{8}{c}}
    \toprule
    &\multicolumn{8}{c}{ESM}\\
     
                        &\multicolumn{1}{c}{6-10am}&\multicolumn{1}{c}{10am-2pm}&\multicolumn{1}{c}{2pm-6pm}&\multicolumn{1}{c}{6pm-10pm}&\multicolumn{1}{c}{6-10am}&\multicolumn{1}{c}{10am-2pm}&\multicolumn{1}{c}{2pm-6pm}&\multicolumn{1}{c}{6pm-10pm}\\
    
                        &\multicolumn{1}{c}{(1)}&\multicolumn{1}{c}{(2)}&\multicolumn{1}{c}{(3)}&\multicolumn{1}{c}{(4)}&\multicolumn{1}{c}{(5)}&\multicolumn{1}{c}{(6)}&\multicolumn{1}{c}{(7)}&\multicolumn{1}{c}{(8)}\\
                        
    \midrule
    
    \textbf{Residential}\\\midrule
Log(Pop. Den)       &       0.010         &      -0.008         &      -0.018         &      -0.055\sym{***}&      -0.014         &       0.000         &      -0.012         &      -0.046\sym{**} \\
                    &     (0.018)         &     (0.018)         &     (0.018)         &     (0.018)         &     (0.017)         &     (0.017)         &     (0.017)         &     (0.017)         \\
Log(MH Income)      &       0.077\sym{***}&       0.091\sym{***}&       0.113\sym{***}&       0.094\sym{***}&       0.087\sym{***}&       0.101\sym{***}&       0.131\sym{***}&       0.120\sym{***}\\
                    &     (0.013)         &     (0.014)         &     (0.014)         &     (0.014)         &     (0.013)         &     (0.014)         &     (0.014)         &     (0.013)         \\
Resi. Div.          &       0.119\sym{***}&       0.166\sym{***}&       0.182\sym{***}&       0.187\sym{***}&       0.120\sym{***}&       0.164\sym{***}&       0.178\sym{***}&       0.178\sym{***}\\
                    &     (0.010)         &     (0.010)         &     (0.010)         &     (0.009)         &     (0.009)         &     (0.009)         &     (0.009)         &     (0.009)         \\
    \textbf{Place}\\\midrule
    Log(Food)           &       0.036\sym{***}&       0.042\sym{***}&       0.043\sym{***}&       0.054\sym{***}&       0.022\sym{**} &       0.016         &       0.021\sym{**} &       0.021\sym{**} \\
                    &     (0.010)         &     (0.010)         &     (0.010)         &     (0.010)         &     (0.010)         &     (0.010)         &     (0.010)         &     (0.009)         \\
Log(Shopping)       &       0.002         &       0.029\sym{***}&       0.032\sym{***}&       0.045\sym{***}&      -0.006         &      -0.020\sym{***}&      -0.025\sym{***}&      -0.006         \\
                    &     (0.007)         &     (0.007)         &     (0.007)         &     (0.007)         &     (0.007)         &     (0.007)         &     (0.007)         &     (0.007)         \\
Log(College)        &       0.020\sym{***}&       0.024\sym{***}&       0.026\sym{***}&       0.014\sym{*}  &       0.009         &       0.014\sym{**} &       0.019\sym{***}&       0.011         \\
                    &     (0.006)         &     (0.007)         &     (0.007)         &     (0.007)         &     (0.006)         &     (0.006)         &     (0.006)         &     (0.007)         \\
Log(Work)           &       0.027\sym{***}&       0.018\sym{**} &       0.011\sym{*}  &       0.012\sym{*}  &       0.020\sym{***}&       0.012\sym{**} &       0.007         &       0.016\sym{**} \\
                    &     (0.007)         &     (0.007)         &     (0.007)         &     (0.007)         &     (0.006)         &     (0.006)         &     (0.006)         &     (0.006)         \\
Log(Sports)         &      -0.003         &      -0.007         &      -0.006         &      -0.007         &      -0.005         &      -0.008         &      -0.009         &      -0.015\sym{**} \\
                    &     (0.006)         &     (0.007)         &     (0.007)         &     (0.007)         &     (0.006)         &     (0.007)         &     (0.007)         &     (0.006)         \\
Log(Transportation) &       0.012\sym{*}  &       0.013\sym{*}  &       0.016\sym{**} &       0.013\sym{*}  &      -0.001         &       0.010         &       0.010         &       0.006         \\
                    &     (0.007)         &     (0.007)         &     (0.007)         &     (0.007)         &     (0.007)         &     (0.007)         &     (0.007)         &     (0.007)         \\
Log(Service)        &      -0.005         &      -0.023\sym{**} &      -0.019\sym{**} &      -0.007         &      -0.008         &      -0.023\sym{**} &      -0.017\sym{**} &      -0.004         \\
                    &     (0.009)         &     (0.008)         &     (0.009)         &     (0.008)         &     (0.008)         &     (0.008)         &     (0.008)         &     (0.008)         \\
Log(Health)         &       0.030\sym{***}&       0.026\sym{***}&       0.033\sym{***}&       0.016\sym{**} &       0.017\sym{**} &       0.015\sym{**} &       0.024\sym{***}&       0.011\sym{*}  \\
                    &     (0.006)         &     (0.006)         &     (0.006)         &     (0.006)         &     (0.006)         &     (0.006)         &     (0.006)         &     (0.006)         \\
Log(Grocery)        &      -0.026\sym{***}&      -0.018\sym{**} &      -0.021\sym{***}&      -0.024\sym{***}&      -0.031\sym{***}&      -0.023\sym{***}&      -0.028\sym{***}&      -0.032\sym{***}\\
                    &     (0.007)         &     (0.007)         &     (0.007)         &     (0.007)         &     (0.007)         &     (0.007)         &     (0.007)         &     (0.007)         \\
Log(Entertainment)  &       0.012         &       0.012         &       0.024\sym{***}&       0.045\sym{***}&       0.021\sym{**} &       0.020\sym{**} &       0.027\sym{***}&       0.030\sym{***}\\
                    &     (0.009)         &     (0.009)         &     (0.008)         &     (0.008)         &     (0.008)         &     (0.008)         &     (0.008)         &     (0.008)         \\
Log(Coffee / Tea)   &       0.038\sym{***}&       0.038\sym{***}&       0.043\sym{***}&       0.040\sym{***}&       0.018\sym{**} &       0.022\sym{***}&       0.027\sym{***}&       0.029\sym{***}\\
                    &     (0.008)         &     (0.008)         &     (0.008)         &     (0.007)         &     (0.007)         &     (0.007)         &     (0.007)         &     (0.007)         \\
Log(Arts / Museum)  &       0.008         &       0.008         &       0.013\sym{*}  &       0.020\sym{***}&       0.003         &       0.003         &       0.006         &       0.011\sym{*}  \\
                    &     (0.007)         &     (0.007)         &     (0.007)         &     (0.007)         &     (0.007)         &     (0.007)         &     (0.007)         &     (0.006)         \\
Log(City / Outdoor) &      -0.005         &      -0.001         &       0.001         &      -0.007         &      -0.012         &      -0.005         &      -0.003         &      -0.009         \\
                    &     (0.008)         &     (0.008)         &     (0.008)         &     (0.008)         &     (0.008)         &     (0.008)         &     (0.008)         &     (0.008)         \\
Street Score        &      -0.025\sym{**} &      -0.001         &      -0.011         &      -0.020\sym{**} &      -0.005         &       0.015         &       0.008         &      -0.000         \\
                    &     (0.010)         &     (0.010)         &     (0.010)         &     (0.010)         &     (0.010)         &     (0.010)         &     (0.010)         &     (0.009)         \\
                        \textbf{Visitor Counts}\\\midrule
Log(Visitors 6 -10)       &                     &                     &                     &                     &       0.404\sym{***}&                     &                     &                     \\
                    &                     &                     &                     &                     &     (0.018)         &                     &                     &                     \\
Log(Visitors 10-14)      &                     &                     &                     &                     &                     &       0.293\sym{***}&                     &                     \\
                    &                     &                     &                     &                     &                     &     (0.014)         &                     &                     \\
Log(Visitors 14-18)      &                     &                     &                     &                     &                     &                     &       0.285\sym{***}&                     \\
                    &                     &                     &                     &                     &                     &                     &     (0.013)         &                     \\
Log(Visitors 18-22)      &                     &                     &                     &                     &                     &                     &                     &       0.312\sym{***}\\
                    &                     &                     &                     &                     &                     &                     &                     &     (0.013)         \\

    \midrule
Observations        &        6289         &        6289         &        6289         &        6289         &        6289         &        6289         &        6289         &        6289         \\
R-squared           &       0.122         &       0.170         &       0.186         &       0.197         &       0.181         &       0.229         &       0.246         &       0.262         \\
    Segment Length  &       YES         &       YES        &       YES        &       YES         &       YES         &       YES        &       YES        &       YES         \\
    
    Fixed effects (county)       &       YES         &       YES        &       YES        &       YES        &       YES         &       YES        &       YES        &       YES         \\
    
    \bottomrule
    \end{tabular}

\label{subtabtimevariantESM}
\smallskip
\scriptsize

    \begin{tablenotes}\centering
    
    \item OLS at street level. Standard errors in parentheses. $\#$ refers to count.\\
    \item $^{***}$ denotes a coefficient significant at the ${0.5\%}$ level, $^{**}$ at the ${5\%}$ level, and $^{*}$ at the ${10\%}$ level. Only street segment with at least 20 unique visitors at each time period are included in the experiment.
    
    \end{tablenotes}

\end{table*}
}

\section{Explain the changes of ESM}
To understand the change of ESM, we collect a series of data from different data sources. The first dataset is from Hidden Boston Restaurant\cite{hiddenboston}. The second data source is ReferenceUSA, which contains information related to business's year of establishment. We only consider those business that are established in 2017. Table \ref{subtabchangeofESM} shows the full results. Column 2 and 4 considers the change of food-related business from Hidden Boston Restaurants. Column 5 only considers the number of new restaurants opened during the study period. Column 7 considers all new business established in 2017 reported by the Reference USA.

\begin{figure*}
    \centering
    \includegraphics{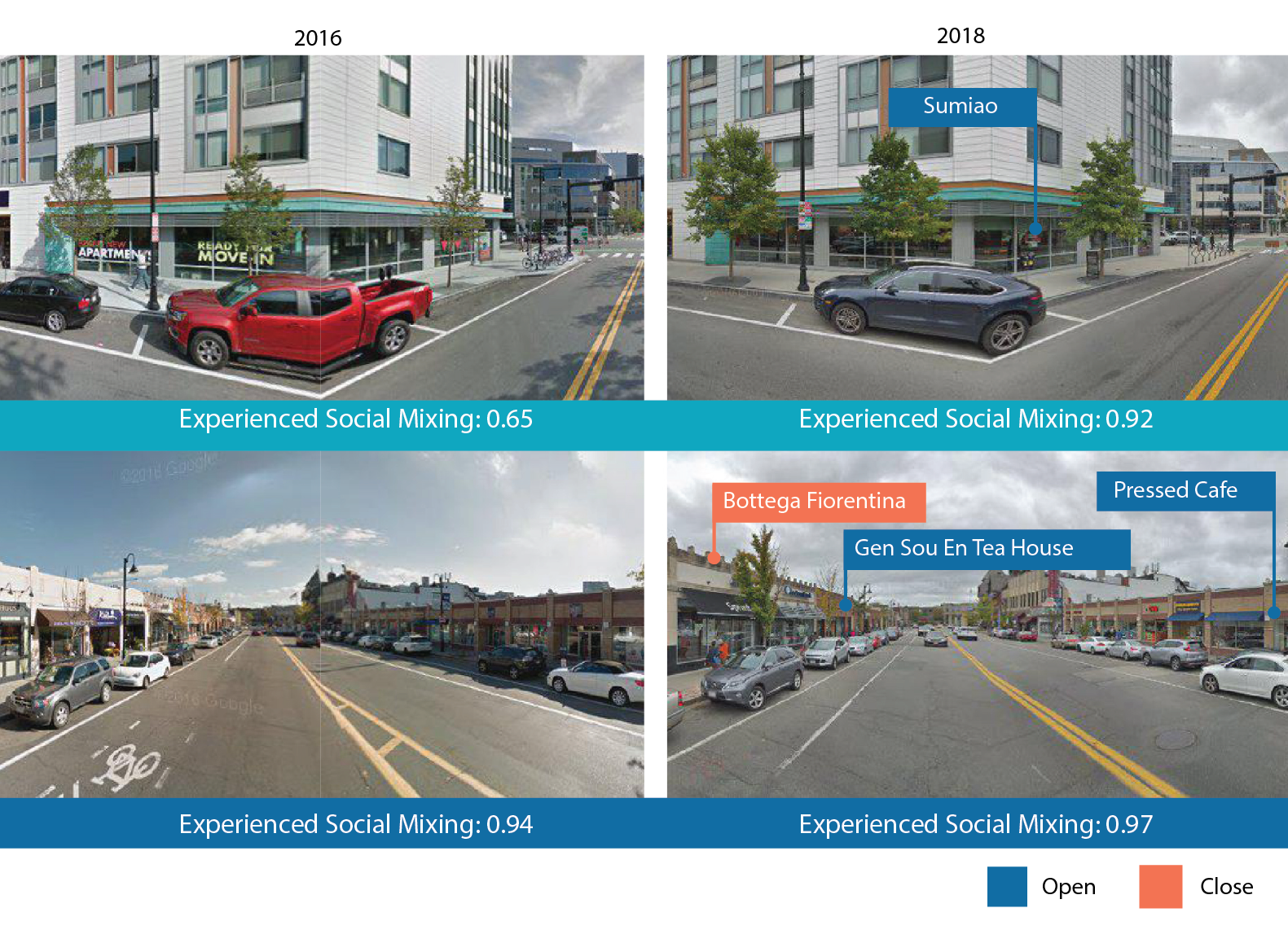}
    \caption{Changes of food-related business identified from Google Street Views}
    \label{sfigchangeGSV}
\end{figure*}

We also verify the changes by randomly sampling from the historical Google Street Views Fig.\ref{sfigchangeGSV}. 

{
\def\sym#1{\ifmmode^{#1}\else\(^{#1}\)\fi}
\begin{table*}[!ht]
    \centering
    \caption{Summary statistics: Boston 2016 - 2018}
    \label{tabstr1}

    \begin{tabular}{l*{3}{cccc}}
    \toprule
                    &\multicolumn{2}{c}{(1)}&\multicolumn{2}{c}{(2)}&\multicolumn{2}{c}{(3)}     \\
                    &\multicolumn{2}{c}{Year 2016}&\multicolumn{2}{c}{Year 2018}&\multicolumn{2}{c}{Difference 2016 - 2018}\\
                    Mean & SD & Mean & SD & $\Delta$ & SE\\
    
    \midrule
    Pop. Den.       &11776.440&10480.827&12296.922&10891.704& 520.482\sym{*}  & (2.127)\\
    MH Income.      &86862.701&25148.994&95737.392&28560.755&8874.691\sym{***}&(14.409)\\
    \% Bachelor     &    0.740&    0.135&    0.752&    0.132&   0.012\sym{***}& (3.893)\\
    Resi. Mixing &    0.700&    0.242&    0.700&    0.235&   0.001         & -0.117)\\
    Street Score    &    4.988&    0.293&    4.997&    0.232&   0.009         & (1.003)\\
    ESM             &    0.805&    0.166&    0.862&    0.124&   0.057\sym{***}&(17.048)\\
    \midrule
    Observations    &     3782&         &     3782&         &     7564         &         \\
    \bottomrule
    \end{tabular}
\smallskip
\scriptsize

    \begin{tablenotes}\centering
    \item Statistics summary from 2016 to 2018 in Boston. Only streets with at least 20 unique visitors in both observation periods (84 days in each year) are included. Each street segment has at least 3 POI within the 100 meter buffer radius. Standard errors in parentheses. $\#$ refers to count. 
    \item $^{***}$ denotes a coefficient significant at the ${0.5\%}$ level, $^{**}$ at the ${5\%}$ level, and $^{*}$ at the ${10\%}$ level.
    
    \end{tablenotes}
\end{table*}
}

{
\def\sym#1{\ifmmode^{#1}\else\(^{#1}\)\fi}
\begin{table*}[!ht]
    \centering
    \caption{Change of ESM 2016 - 2018 (Street-Level)}
    
    \scriptsize
    
    \begin{tabular}{l*{7}{c}}
    \toprule
    &\multicolumn{7}{c}{$\Delta$ ESM}\\
                        &\multicolumn{1}{c}{(1)}&\multicolumn{1}{c}{(2)}&\multicolumn{1}{c}{(3)}&\multicolumn{1}{c}{(4)}&\multicolumn{1}{c}{(5)}&\multicolumn{1}{c}{(6)}&\multicolumn{1}{c}{(7)}\\
    
    \midrule
    \textbf{Trend Controls}\\
    Segment Length      &       0.012\sym{***}&       0.011\sym{***}&       0.012\sym{***}&       0.012\sym{***}&       0.011\sym{***}&       0.012\sym{***}&       0.011\sym{***}\\
                    &     (0.002)         &     (0.002)         &     (0.003)         &     (0.002)         &     (0.002)         &     (0.002)         &     (0.002)         \\
Log(Pop. Den)       &       0.006\sym{**} &       0.006\sym{**} &       0.004         &       0.006\sym{**} &       0.008\sym{***}&       0.006\sym{**} &       0.006\sym{**} \\
                    &     (0.003)         &     (0.003)         &     (0.005)         &     (0.003)         &     (0.003)         &     (0.003)         &     (0.003)         \\
Log($\#$ POI)       &       0.016\sym{***}&       0.015\sym{***}&       0.017\sym{***}&       0.016\sym{***}&       0.015\sym{***}&       0.014\sym{***}&       0.015\sym{***}\\
                    &     (0.002)         &     (0.002)         &     (0.003)         &     (0.002)         &     (0.002)         &     (0.002)         &     (0.002)         \\
ESM. 2016           &      -0.128\sym{***}&      -0.127\sym{***}&      -0.125\sym{***}&      -0.128\sym{***}&      -0.127\sym{***}&      -0.128\sym{***}&      -0.128\sym{***}\\
                    &     (0.002)         &     (0.002)         &     (0.004)         &     (0.002)         &     (0.002)         &     (0.002)         &     (0.002)         \\
                        
                        \midrule
    \textbf{Residential Features}\\
    $\Delta$ Pop Den    &       0.003         &                     &       0.013\sym{**} &       0.003         &       0.003\sym{*}  &       0.003         &       0.003         \\
                    &     (0.002)         &                     &     (0.005)         &     (0.002)         &     (0.002)         &     (0.002)         &     (0.002)         \\
$\Delta$ MH Income  &      -0.000         &                     &       0.002         &      -0.000         &      -0.000         &      -0.000         &      -0.000         \\
                    &     (0.002)         &                     &     (0.004)         &     (0.002)         &     (0.002)         &     (0.002)         &     (0.002)         \\
$\Delta \%$ Bachelor&       0.010\sym{***}&                     &       0.008\sym{*}  &       0.010\sym{***}&       0.009\sym{***}&       0.010\sym{***}&       0.010\sym{***}\\
                    &     (0.002)         &                     &     (0.004)         &     (0.002)         &     (0.002)         &     (0.002)         &     (0.002)         \\
$\Delta$ Resi. Diversity&      -0.001         &                     &       0.007\sym{*}  &      -0.001         &      -0.001         &      -0.001         &      -0.001         \\
                    &     (0.002)         &                     &     (0.004)         &     (0.002)         &     (0.002)         &     (0.002)         &     (0.002)         \\

    \midrule
    \textbf{Built Environment}\\
    $\Delta$ Food Business &                     &       0.004\sym{**} &       0.004\sym{**} &       0.006\sym{***}&       0.003\sym{**} &                     &                     \\
                    &                     &     (0.001)         &     (0.002)         &     (0.002)         &     (0.001)         &                     &                     \\
    $\Delta$ Street Score&                     &                     &       0.026         &                     &                     &                     &                     \\
                    &                     &                     &     (0.062)         &                     &                     &                     &                     \\
$\Delta$ Street Score (sq)&                     &                     &      -0.025         &                     &                     &                     &                     \\
                    &                     &                     &     (0.062)         &                     &                     &                     &                     \\
    $\Delta$ Food Business \\
    $\times$ ESM. 2016&                     &                     &                     &      -0.006\sym{**} &                     &                     &                     \\
                    &                     &                     &                     &     (0.003)         &                     &                     &                     \\
    $\#$ New Food Business          &                     &                     &                     &                     &                     &       0.008\sym{***}&                     \\
                    &                     &                     &                     &                     &                     &     (0.001)         &                     \\
    $\Delta$ New Business&                     &                     &                     &                     &                     &                     &       0.005\sym{***}\\
                    &                     &                     &                     &                     &                     &                     &     (0.001)         \\
                        
    $\Delta$ Log(Visitors)&                     &                     &                     &                     &       0.017\sym{***}&                     &                     \\
                    &                     &                     &                     &                     &     (0.002)         &                     &                     \\
    \midrule
    Observations        &        3768         &        3768         &        1417         &        3768         &        3768         &        3768         &        3768         \\
R-squared           &      0.5480         &      0.5450         &      0.5272         &      0.5490         &      0.5579         &      0.5499         &      0.5489         \\
    Fixed effects \\
    (county) & Yes & Yes & Yes & Yes & Yes & Yes & Yes\\

    \bottomrule
    \end{tabular}

\label{subtabchangeofESM}
\scriptsize

    \begin{tablenotes}\centering
    \item OLS estimates on change of ESM from 2016 to 2018. Only streets with at least 30 unique visitors in both observation periods (84 days in each year) are included. Standard errors in parentheses. $\#$ refers to count.  Column 4 - 6 uses food business data reported from HiddenBoston. Column 7 use data from Reference USA 2018 by extracting the businesses that are established in 2017. $\#$New Food refers to new food business that opened from Jan. 2017 to September 2018. $\Delta$ Food refers to number of new food business subtracting by closed food business from Jan. 2017 to September 2018.
    
    \item $^{***}$ denotes a coefficient significant at the ${0.5\%}$ level, $^{**}$ at the ${5\%}$ level, and $^{*}$ at the ${10\%}$ level.
    \end{tablenotes}

\end{table*}
}

\section{Safe Score}
\begin{figure*}
    \centering
    \includegraphics{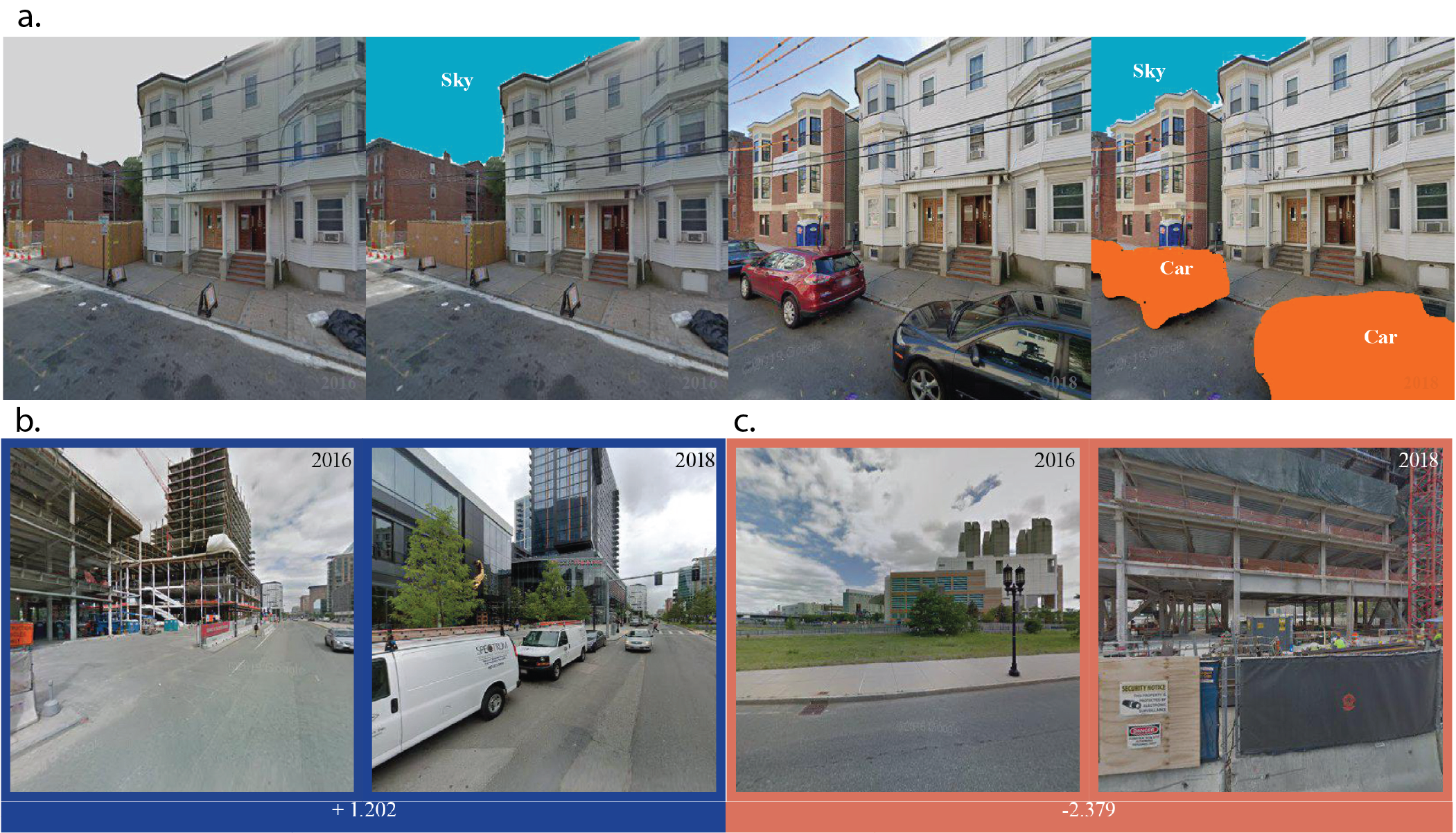}
    \caption{Processing the Google Street View to predict safety score. a. We first use the segmentation model to detect sky, cars, and people and mask them out from the image with an white mask. Then we predict the images' safety score. b. Examples of changes of street scores between 2016 and 2018.}
    \label{sfigGSVsafety}
\end{figure*}

Fig. \ref{sfigGSVsafety} shows the process we adopt to predict the safety score. It also shows examples of changes of safety score identified from 2016 to 2018.

\section{POI venues}
Table \ref{tabpoisummary} summarizes the overall distribution of POI venues we downloaded from Foursquares and used in this study. Detail categories of each POI was the same as used in Moro et al. \cite{moro2021mobility}

\begin{table}[!ht]
\centering
\caption{Summary of POI categories included in the study}
\label{tabpoisummary}
\begin{tabular}{@{}llll@{}}

\toprule
city            & Boston & New York & Philadelphia \\ \midrule
Arts / Museum   & 315    & 1635    & 482          \\
City / Outdoors & 233    & 2077    & 313          \\
Coffee / Tea    & 356    & 1486    & 455          \\
College         & 527    & 1143    & 464          \\
Entertainment   & 553    & 3602    & 807          \\
Food            & 3450   & 15918   & 4256         \\
Grocery         & 809    & 3616    & 902          \\
Health          & 470    & 2067    & 746          \\
Residential     & 0      & 20      & 2            \\
Service         & 4170   & 18799   & 5595         \\
Shopping        & 2555   & 10081   & 3676         \\
Sports          & 128    & 497     & 150          \\
Transportation  & 190    & 994     & 406          \\
Work            & 415    & 1900    & 504          \\ \midrule
Total POI       & 14410  & 64548   & 19060        \\ \bottomrule
\end{tabular}
\end{table}

\newpage
\section{Crime and ESM}
Crime reports are downloaded from: 
\begin{itemize}
    \item https://www1.nyc.gov/,
    \item https://data.boston.gov/dataset/crime-incident-reports-august-2015-to-date-source-new-system,
    \item https://www.opendataphilly.org/dataset/crime-incidents,
    \item https://data.cambridgema.gov/Public-Safety/Crime-Reports/xuad-73uj.
\end{itemize}
 
We estimate the relationship between crime and ESM by regressing ESM on street-level aggregated crime counts:
\begin{equation}
    Log(Crimes) = \beta ESM + \alpha X,
\end{equation}
where we separated crimes into violent crimes and pretty crimes. The violent crimes include aggregated assaults, rape, murder, and robbery. The pretty crimes include theft-related crimes. $X$ includes number of visitors, count of POIs, population density, median household income, residential diversity, street segment length, and county-level fixed effects.
Table \ref{tabcrime} report all results. Conditioning on the number of visitors, POI, population density, median household income, the ESM has a negative relationship with the crime count. On the contrary, residential diversity positively correlates with the crime count. Note that crime study is only constrained within each main city boundary given the data availability. The rest of the study is conducted at the metropolitan level. 

\begin{figure*}
    \centering
    \includegraphics[width = 0.9\textwidth]{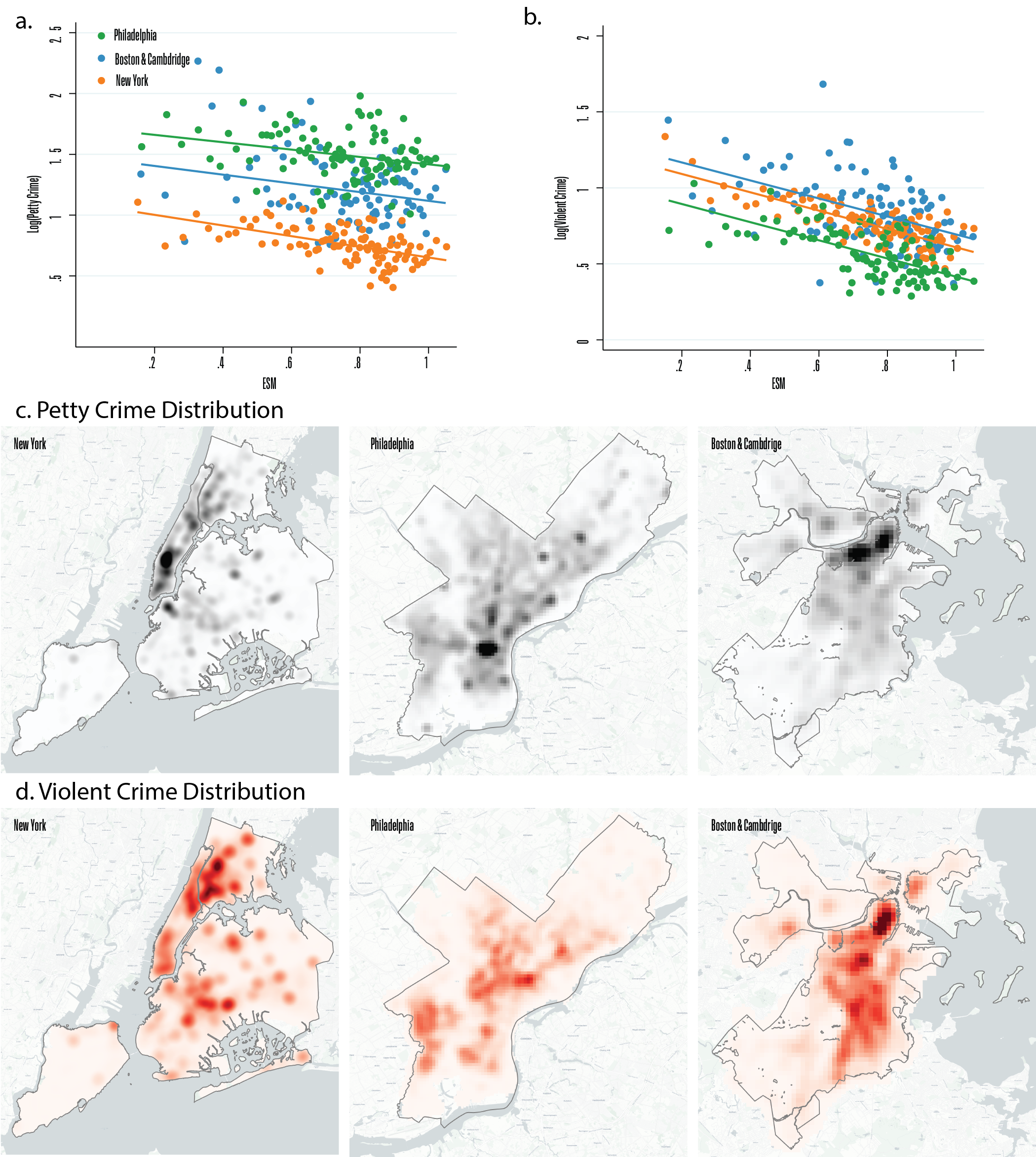}
    \caption{a. Binned Scatter plot between ESM and Log-transformed pretty crime counts. b. Binned Scatter plot between ESM and Log-transformed violent crime counts. c. Pretty crime heatmap within each city's boundary. d. Violent crime heatmap within each city's boundary.}
    \label{sfigcrime}
\end{figure*}
{
\begin{table*}[!htb]
    \centering
\scriptsize
    \caption{Crimes and Street ESM}

{
\def\sym#1{\ifmmode^{#1}\else\(^{#1}\)\fi}
\begin{tabular}{l*{8}{c}}
\toprule
                    &\multicolumn{1}{c}{(1)}&\multicolumn{1}{c}{(2)}&\multicolumn{1}{c}{(3)}&\multicolumn{1}{c}{(4)}&\multicolumn{1}{c}{(5)}&\multicolumn{1}{c}{(6)}&\multicolumn{1}{c}{(7)}&\multicolumn{1}{c}{(8)}\\
                    &\multicolumn{4}{c}{Log(Violent Crime Count)}&\multicolumn{4}{c}{Log(Petty Crime Count)}\\

\midrule
ESM &      -0.037\sym{***}&      -0.098\sym{***}&      -0.103\sym{***}&      -0.024\sym{***}&       0.062\sym{***}&      -0.071\sym{***}&      -0.090\sym{***}&      -0.026\sym{**} \\
                    &     (0.008)         &     (0.008)         &     (0.008)         &     (0.008)         &     (0.010)         &     (0.010)         &     (0.010)         &     (0.010)         \\

Log(Visitors)       &                     &       0.207\sym{***}&       0.193\sym{***}&       0.180\sym{***}&                     &       0.454\sym{***}&       0.405\sym{***}&       0.397\sym{***}\\
                    &                     &     (0.008)         &     (0.009)         &     (0.009)         &                     &     (0.013)         &     (0.014)         &     (0.013)         \\
Log(\#POI)          &                     &                     &       0.040\sym{***}&       0.115\sym{***}&                     &                     &       0.139\sym{***}&       0.201\sym{***}\\
                    &                     &                     &     (0.009)         &     (0.009)         &                     &                     &     (0.013)         &     (0.013)         \\
Log(Pop. Den)       &                     &                     &                     &       0.147\sym{***}&                     &                     &                     &       0.145\sym{***}\\
                    &                     &                     &                     &     (0.011)         &                     &                     &                     &     (0.015)         \\
Log(MH Income)      &                     &                     &                     &      -0.338\sym{***}&                     &                     &                     &      -0.298\sym{***}\\
                    &                     &                     &                     &     (0.010)         &                     &                     &                     &     (0.015)         \\
Resi. Div.          &                     &                     &                     &       0.028\sym{***}&                     &                     &                     &       0.050\sym{***}\\
                    &                     &                     &                     &     (0.008)         &                     &                     &                     &     (0.011)         \\
\midrule
Observations        &       12366         &       12366         &       12366         &       12366         &       12366         &       12366         &       12366         &       12366         \\
R-squared           &      0.0610         &      0.1102         &      0.1116         &      0.2163         &      0.0709         &      0.1902         &      0.1989         &      0.2405         \\
Fixed Effect (County) &      Yes         &       Yes         &      Yes        &       Yes         &       Yes         &       Yes         &      Yes         &       Yes         \\
Street Length Control &      Yes         &       Yes         &      Yes        &       Yes         &       Yes         &       Yes         &      Yes         &       Yes         \\
\bottomrule
\end{tabular}
}

    \label{tabcrime}
    \scriptsize

    \begin{tablenotes}\centering
    \item OLS estimates crime count (2016 data) and ESM. Only New York City, Boston, Cambridge, and Philadelphia City are included in the study. Violent crime includes murder, robbery, rape, aggravated assault. Petty crime includes shoplifting, theft, and larceny.
    
    \item $^{***}$ denotes a coefficient significant at the ${0.5\%}$ level, $^{**}$ at the ${5\%}$ level, and $^{*}$ at the ${10\%}$ level.
    \end{tablenotes}
\end{table*}
}

\section{Software}
Analysis was conducted using Python, Jupyter Lab, and Stata. Map graphics were produced using QGIS.
\bibliographystyle{plain}